\documentclass[fleqn,usenatbib]{mnras}

\usepackage[T1]{fontenc}
\usepackage{ae,aecompl}

\usepackage{graphicx}	
\usepackage{graphics} 
\usepackage{amsmath}	
\usepackage{amssymb}	
\usepackage{slashed}    
\usepackage{subcaption}
\usepackage[export]{adjustbox}
\DeclareMathOperator*{\argminT}{argmin}
\DeclareMathOperator*{\argmaxT}{argmax}

\DeclareMathOperator*{\maxT}{max}
\usepackage{mathtools}
\DeclarePairedDelimiterX{\norm}[1]{\lVert}{\rVert}{#1}
\makeatletter
\newcommand*{\rom}[1]{\expandafter\@slowromancap\romannumeral #1@}
\makeatother
\usepackage{wrapfig}
\usepackage{bm}
\usepackage{multirow}
\usepackage{arydshln}
\usepackage{color, colortbl}
\definecolor{LightCyan}{rgb}{0.88,1,1}

\usepackage{pifont}
\newcommand{\cmark}{\ding{51}}%
\newcommand{\xmark}{\ding{55}}%

\usepackage{smartdiagram}
\usepackage{tikz}
\usetikzlibrary{shapes,arrows}
\tikzstyle{block} = [rectangle, draw, fill=yellow!10,
    text width=12em, text centered, rounded corners, minimum height=3em]
\tikzstyle{line} = [draw, -latex']
\tikzstyle{decision} = [rectangle, draw, fill=green!10, text width=12em, text centered, rounded corners, minimum height=3em]
\tikzstyle{parameter_fixed} = [rectangle, draw, fill=purple!10, text width=12em, text centered, rounded corners, minimum height=3em]
\tikzstyle{iteration} = [rectangle, draw, fill=blue!10, text width=12em, text centered, rounded corners, minimum height=3em]

\usetikzlibrary{decorations.pathreplacing}

\usepackage{xcolor}
\usepackage{soul}

\newcommand{\emdash}{\nobreak--\nobreak\hskip4pt}

\usepackage{algorithm}
\usepackage{setspace}
\usepackage[noend]{algpseudocode}
\makeatletter
\def\BState{\State\hskip-\ALG@thistlm}
\makeatother

\title[Sparse Bayesian mass-mapping on the celestial sphere]{Sparse Bayesian mass-mapping with uncertainties: \\
full sky observations on the celestial sphere}

\author[Price et al.]{
M.~A.~Price$^{1}$\thanks{E-mail: m.price.17@ucl.ac.uk}, J.~D.~McEwen$^{1}$, L.~Pratley$^{1,2 }$, and T.~D.~Kitching$^{1}$ \\
$^{1}$Mullard Space Science Laboratory, University College London, RH5 6NT, UK\\
$^{2}$Dunlap Institute for Astronomy and Astrophysics, University of Toronto, ON M5S 3H4, Canada\\
}

\date{Accepted XXX. Received YYY; in original form ZZZ}

\pubyear{2018}

\begin{document}
\label{firstpage}
\pagerange{\pageref{firstpage}--\pageref{lastpage}}
\maketitle

\begin{abstract}
To date weak gravitational lensing surveys have typically been restricted to small fields of view, such
that the \textit{flat-sky approximation} has been sufficiently satisfied. However, with Stage \rom{4}
surveys (\textit{e.g. LSST} and \textit{Euclid}) imminent, extending mass-mapping
techniques to the sphere is a fundamental necessity. As such, we extend
the sparse hierarchical Bayesian mass-mapping formalism presented in previous work to the
spherical sky. For the first time, this allows us to construct \textit{maximum a posteriori} spherical
weak lensing dark-matter mass-maps, with principled Bayesian uncertainties, without imposing
or assuming Gaussianty. We solve the spherical mass-mapping inverse
problem in the analysis setting adopting a
sparsity promoting Laplace-type wavelet prior, though this theoretical framework supports
all log-concave posteriors.
Our spherical mass-mapping formalism facilitates principled statistical interpretation of
reconstructions. We apply our framework to convergence reconstruction on high resolution
N-body simulations with pseudo-Euclid masking, polluted with a variety of realistic noise levels,
and show a significant increase in reconstruction fidelity compared to standard approaches.
Furthermore we perform the largest joint reconstruction to date of the majority of publicly available shear
observational datasets (combining DESY1, KiDS450 and CFHTLens) and find that our formalism recovers
a convergence map with significantly enhanced small-scale detail. Within
our Bayesian framework we validate, in a statistically rigorous manner, the community's intuition
regarding the need to smooth spherical Kaiser-Squires estimates to provide physically meaningful convergence maps.
Such approaches cannot reveal the small-scale physical structures that we recover within our
framework.
\end{abstract}

\begin{keywords}
gravitational lensing: weak -- (\textit{Cosmology:}) large-scale structure of Universe -- Methods: statistical -- Methods: data analysis -- techniques: image processing -- techniques: compressed sensing
\end{keywords}



\section{Introduction} \label{sec:introduction}
Gravitational lensing is an astrophysical phenomenon through which the geometry of distant
galaxies becomes distorted by the intervening matter distribution. Mathematically, this lensing
effect is a perturbation by the local matter topology of the null geodesics along which photons
travel \citep{[59],[1],[2]}. As such, gravitational lensing is sensitive
to all matter (both visible and invisible) and is thus a natural tool with which to probe the
nature of dark matter.
\par
Weak gravitational lensing refers to the vast majority of lensing events for which
images are not multiply sourced or `strongly lensed'. Equivalently the weak lensing regime
can be defined as the regime
in which the lensing perturbations remain (to a good approximation) linear. At first order the
effect of weak lensing on distant galaxy images manifests itself as two quantities: the spin-0
magnification referred to as the \textit{convergence} field ${}_0\kappa$, and a spin-2
perturbation to the ellipticity (third-flattening) referred to as the \textit{shearing} or \textit{shear}
field ${}_2\gamma$.
\par
Due to the \textit{`mass-sheet degeneracy'} there is no way to construct \textit{a priori}
estimates of the intrinsic brightness, hence the convergence field is an unobservable
quantity -- theoretically one could infer the convergence field directly from the galaxy sizes,
but the intrinsic dispersion is too high \citep{Alsing_2015_mags}. However, as the distribution of instrinsic ellipticities
has zero mean and sufficiently tight dispersion, averaging sufficient observations within a
given pixel can provide an accurate estimator for the shear signal. As such, measurements
of the shear field ${}_2\gamma$ are typically taken and inverted to form estimates of
${}_0\kappa$ \emdash coined \textit{dark matter mass maps} by \citet{[25]}.
\par
A large proportion \citep{Taylor_2019} of cosmological information can be extracted directly
from the shear field \citep{Waerbeke_2013, Fluri_2019_NN_estimators, Giblin_2018}, however recently
cosmologists have become increasingly interested in extracting information from higher order
statistics, such as peak \& void statistics and Minkowski functionals, which are typically calculated
directly from the convergence field \citep[\textit{e.g.}][]{[27],[28],[60],[61]} \emdash motivating
research into optimal mass-mapping techniques. Typically, these higher order statistics aim
to probe the non-Gaussian information content of the convergence field.
\par
Mapping from shear to convergence (mass-mapping) requires solving an (often seriously)
ill-posed inverse problem \emdash mass-mapping takes the form of a typical noisy
deconvolution problem with a spin-2 kernel  \citep{[3]}, which is classically ill-posed.
The most naive mass-mapping
technique for small fields of view is planar Kaiser-Squires \citep[KS; ][]{[5]} which is direct
inversion of the forward model in Fourier space. This estimator does not take into account
noise or boundary effects, and so is typically post-processed \textit{via} convolution with a
large Gaussian smoothing kernel, thus heavily degrading the quality of
high-resolution non-Gaussian information. Moreover, decomposition of spin-fields on
bounded manifolds is known to be degenerate \citep{[4]} and so for non-trivial
masking the KS estimator is ill-defined and can be shown to perform poorly (see
Section \ref{sec:Sim_testing}).
\par
Many, perhaps more sophisticated, approaches to mass-mapping on the plane have been
developed \citep[\textit{e.g.}][]{[29],[6],[15],[63]} though all either lack a principled
statistical framework or rely heavily on assumptions or impositions of Gaussianity. In previous
work we present a sparse hierarchical Bayesian formalism for planar mass-mapping
\citep{M1, M2, M3} that provides fully principled statistical uncertainties without the need to
assume Gaussianity and without the computational overhead of MCMC methods
\citep[\textit{e.g.}][]{[56],[57],[58]}.
\par
One key assumption of these `planar' mass-mapping techniques is that the area of interest
on the sky can be well approximated as a plane. This assumption is colloquially referred to
as the the \textit{flat-sky approximation}. For small-field surveys this approximation is
typically justified. However for future wide-field Stage \rom{4} surveys mass-mapping
must be constructed natively on the sphere \citep{DES_2018} to avoid errors due to projection
effects, which can be large \citep{[3], Zoe2019}. Naturally one can naively invert the spherical
forward model to form the \textit{spherical Kaiser-Squires} estimator \citep[SKS; ][]{[3]} which
avoids projection effects but is seriously ill-posed, as is the KS method. It should
be noted that alternative techniques for spherical reconstruction have also been developed \citep[\textit{e.g.}][]{[64]}.
\par
In this paper we extend the previously developed hierarchical Bayesian-sparse formalism
to the sphere which, for the first time, allows \textit{maximum a posteriori} (MAP)
convergence reconstruction with principled Bayesian uncertainties in very
high-dimensions natively on the sphere without making any assumptions or impositions of
Gaussianity. Throughout this paper we refer to our estimator, formed within this framework,
as the DarkMapper estimator (and by extension the DarkMapper codebase). The
reconstruction formalism presented in this paper and any uncerainty quantification techniques
that follow support any choice of likelihood or prior such that the posterior function belongs
to the (rather comprehensive) set of log-concave functions. As such one can incorporate
various experimental or systematic effects in future, \textit{e.g.} more complex noise models
or intrinsic alignment corrections \textit{etc.}
\par
The structure of this paper is as follows. In Section \ref{sec:background} we provide
background mathematical details relevant to the scope of this paper, such as the analysis
of spin signals on the sphere, and succiently review weak gravitational lensing.
Following this Section \ref{sec:bayesian_inference} provides a cursory introduction to
Bayesian analysis before presenting and discussing both the general hierarchical Bayesian
formalism and our DarkMapper estimator. In this section we explicitly outline the likelihood
and priors used throughout this paper but place emphasis on the generality of this formalism.
Furthermore, we outline how to fold uncertainty in regularization parameters into the hierarchy
\textit{via} the allocation of a suitable (here a conjugate) hyper-prior distribution.
In Section \ref{sec:UQ} we extend previously developed uncertaintiy quantification
techniques to the spherical space and discuss how one should
approach constructing custom uncertainty quantification techniques which fit within our
formalism. In sections \ref{sec:Sim_testing}, using high resolution N-body \citep{Ryuichi} simulations,
pseudo-Euclid masking (a masking of the galactic plane and the
ecliptic) and noise realisations representative of a variety of weak lensing survey eras (
including Stage \rom{4}) we demonstrate the drastic increase in reconstruction fidelity of
DarkMapper over SKS. Penultimately, in Section \ref{sec:public_data} we apply
both the SKS and DarkMapper estimators to a global weak lensing dataset
constructed \textit{via} the concatenation of the majority of publicly available observational datasets. To
the best of our knowledge this is the first such global spherical dark-matter mass-maps.
Furthermore, we perform global Bayesian uncertainty quantification on these reconstructions.
Finally, in Section \ref{sec:conclusions} we draw conclusions.

\section{Background} \label{sec:background}
Here we present a cursory synopsis of the relevant background required to understand
weak lensing on the sphere. In no way is this a complete description and so we
recommend the reader follow related papers \citep{[40],[3],[42]}.

\subsection{Spin-s Spherical Fields}
Local rotations by $\chi \in [0, 2\pi)$ about the tangent plane centered on the spherical
coordinate $\omega = (\theta, \psi) \in \mathbb{S}^2$ of square integrable spin-s fields for
$s \in \mathbb{Z}$ are defined generally by \citep{[43],[44],[3],[41]}
\begin{equation}
{}_sf^{\prime}(\omega) = e^{-is\chi} {}_sf(\omega),
\end{equation}
where $\omega = (\theta,\psi)$ are standard spherical coordinates, given by co-latitude
$\theta \in [0, \pi)$ and longitude $\psi \in [0,2\pi)$. The natural set of orthogonal basis
functions for spherical fields are the \textit{spherical harmonics} $Y_{\ell m}(\omega)$.
\par
When considering spin-s fields on $\mathbb{S}^2$ the natural set of orthogonal basis
functions are the \textit{spin-weighted spherical harmonics}. The spin weighted spherical
harmonics are generated by application of the spin raising and lowering operators ($\eth$
and $\bar{\eth}$ respectively) to the spherical eigenfunctions $Y_{\ell m}(\omega)$. The spin-s
raising and lowering operators are given respectively, by
\begin{align}
\eth \equiv -\sin^s\theta \Big ( \frac{\partial}{\partial \theta} + \frac{i \partial}{\sin\theta \partial \psi} \Big ) \sin^{-s}\theta, \\
\bar{\eth} \equiv -\sin^{-s}\theta \Big ( \frac{\partial}{\partial \theta} - \frac{i \partial}{\sin\theta \partial \psi} \Big ) \sin^{s}\theta.
\end{align}
\par
On application to ${}_sY_{\ell m}(\omega)$ we find the recursion relations,
\begin{align}
\eth \; {}_sY_{\ell m} (\omega) &= \big [ (\ell - s)(\ell + s +1) \big ]^{1/2} \; {}_{s+1} Y_{\ell m}(\omega), \\
\bar{\eth} \; {}_sY_{\ell m} (\omega) &= -\big [ (\ell + s)(\ell - s +1) \big ]^{1/2} \; {}_{s-1} Y_{\ell m}(\omega).
\end{align}
\par
Following these recursions it is clear that any spin-s weighted spherical harmonic can be
represented as $s \in \mathbb{N}$ repeated applications of the spin raising (lowering) operator $\eth$ to
the standard spin-0 spherical harmonic $Y_{\ell m}$ such that,
\begin{equation}
{}_sY_{\ell m}(\omega) = \Big [ \frac{(\ell - s)!}{(\ell + s)!} \Big ]^{\frac{1}{2}} \; \eth^sY_{\ell m}(\omega),
\end{equation}
for positive semi-definite spin $0 \leq s \leq \ell$, and for negative semi-definite spin
$-\ell \leq s \leq 0$ by,
\begin{equation}
{}_sY_{\ell m}(\omega) = (-1)^s \Big [ \frac{(\ell + s)!}{(\ell - s)!} \Big ]^{\frac{1}{2}} \; \bar{\eth}^{-s}Y_{\ell m}(\omega).
\end{equation}
\par
The spin-s weighted spherical harmonics form a complete set of orthogonal basis functions
which leads to the harmonic representation of a spin-s field ${}_sf(\omega)$ by
\begin{equation} \label{eq:projection_into_spin_harmonics}
{}_s f (\omega) = \sum^{\infty}_{\ell = 0} \sum^{\ell}_{m=-\ell} {}_s\hat{f}_{\ell m} \; {}_sY_{\ell m}(\omega).
\end{equation}
\par
We can then trivially invert this decomposition to give the spin-s field ${}_sf(\omega)$
projected onto the spin basis eigenfunctions (\textit{i.e.} the spin-spherical harmonic
coefficients),
\begin{equation} \label{eq:spin_integral}
{}_s \hat{f}_{\ell m} = \int_{\mathbb{S}^2} d\Omega(\omega) \; {}_sf(\omega) \; {}_sY_{\ell m}^{*}(\omega),
\end{equation}
where the integral is over the sphere $\mathbb{S}^2$, and $d\Omega(\omega) = \sin \theta d\theta d\phi$
is the rotation invariant measure on the sphere. Typically the signal is band-limited at
$\ell_{\text{max}}$ which implies ${}_sf_{\ell m} = 0, \forall \ell \geq \ell_{\text{max}}$ allowing
the $\ell$ summations in equation (\ref{eq:projection_into_spin_harmonics}) and the upper limit of the integral
in equation (\ref{eq:spin_integral}) to be truncated at $\ell_{\text{max}}$ to make the computation
tractable.

\subsection{Weak Lensing on the Sphere}
This section provides a basic introduction to weak lensing mass-mapping in the spherical
setting. For a more detailed introduction, we refer the reader to popular reviews
\citep[\textit{e.g.}][]{[1],[2]}.
\par
Gravitational lensing is an astrophysical effect which describes the deflection of distant
photons as they propagate to us here and now by the intervening local matter distribution.
As lensing is sensitive to the local matter distribution (both visible and dark), it provides a
natural cosmological probe of dark matter.
\par
Specifically, the weak lensing (WL) regime refers to photons which have angular position
on the source plane $\beta$ (relative to the line-of-sight from observer through the lensing
mass) smaller than one Einstein radius $\theta_E$ to the intervening lensing mass.
Mathematically this restricts us to singular solutions of the lens equation,
\begin{equation}
\beta = \theta - \theta_E^2 \frac{\theta}{|\theta|^2}, \quad \text{where} \quad \theta_E = \sqrt{\frac{4GM}{c^2}\frac{f_K(r - r^{\prime})}{f_K(r)f_K(r^{\prime})}},
\end{equation}
for angular diameter distance $f_K$, defined in the usual sense, which is dependent on
the curvature of the Universe $K$. The Universe has been observed to be essentially flat
\citep{planck2018} and so to a good approximate $K \approx 0 \Rightarrow f_K(r) \approx r$,
where $r$ is the comoving distance.
\par
Galaxies are naturally sparsely distributed across the sky and so the overwhelming
majority of observations fall within the weak lensing regime \citep{[1]}. Now consider a
lensing potential $\phi$ which is the weighted integral along the line of sight of the local
Newtonian potential $\Phi$,
\begin{equation}
\phi(r,\omega) = \frac{2}{c^2} \int_0^r dr^{\prime} \frac{f_K(r - r^{\prime})}{f_K(r) f_K(r^{\prime})} \Phi(r^{\prime}, \omega).
\end{equation}
Poisson's equation must then be satisfied by the local Newtonian potential,
\begin{equation}
\nabla^2 \Phi(r,\omega) = \frac{3 \Omega_M H_0^2}{2a(r)} \delta(r,\omega),
\end{equation}
where $\delta(r,\omega)$ is the fractional over-density, $H_0$ is the Hubble constant,
$a(r)$ is the scale-parameter and $\Omega_M$ is the matter density parameter. At first
order two physical lensing quantities can be constructed, these being the gravitational
shear ${}_2\gamma$ and the convergence ${}_0\kappa$ \citep{[1],[2]}, where the
subscripts reflect the spin of each field.
\par
These quantities are related to the underlying scalar integrated potential ${}_0\phi$ by the
relations \citep{Castro_2005, [3]},
\begin{align}
& _0\kappa(r,\omega) = \frac{1}{4}(\eth \bar{\eth} + \bar{\eth} \eth) \; {}_0\phi(r,\omega), \label{eq:kappatophi} \\
& _2\gamma(r,\omega) = \frac{1}{2} \eth \eth \; {}_0\phi(r,\omega), \label{eq:gammatophi}
\end{align}
If we now project these values into their harmonic representations by equation (\ref{eq:spin_integral})
we find the harmonic space relations,
\begin{align}
{}_0\hat{\kappa}_{\ell m} &= -\frac{1}{2}\ell (\ell + 1) \; {}_0\hat{\phi}_{\ell m}, \\
{}_2\hat{\gamma}_{\ell m} &= \frac{1}{2} \sqrt{ \frac{(\ell + 2)!}{(\ell - 2)!} } \; {}_0\hat{\phi}_{\ell m}.
\end{align}
We can then trivially draw a relationship between ${}_2\hat{\gamma}_{\ell m}$ and ${}_0\hat{\kappa}_{\ell m}$in harmonic space,
\begin{equation}
{}_2\hat{\gamma}_{\ell m} = \mathcal{W}_{\ell} \; {}_0\hat{\kappa}_{\ell m},
\end{equation}
which is the \textit{spherical forward model}. We have defined a mapping kernel
\citep[as in \textit{e.g.}][]{[3]} in harmonic space such that,
\begin{equation} \label{eq:spherical_kernel}
\mathcal{W}_{\ell} = \frac{-1}{\ell (\ell + 1)}\sqrt{ \frac{(\ell + 2)!}{(\ell - 2)!} }.
\end{equation}
This mapping is analogous to the planar forward model \citep{M1} but now defined on
$\mathbb{S}^2$. This mapping can trivially be inverted to define the so-called `Spherical
Kaiser-Squires' \citep[SKS,][]{[3]} convergence estimator,
\begin{equation}
{}_0\hat{\kappa}_{\ell m}^{\text{SKS}} = \mathcal{W}_{\ell}^{-1} \; {}_2\hat{\gamma}_{\ell m}^{\text{obs}},
\end{equation}
where superscript `obs' refers to the observations (or measurements) of a given shear field
${}_2\gamma$. A real-space representation of this mapping exists \citep{[3]}.

It is of interest to notice certain similarities between the SKS estimator ${}_0\kappa_{\text{SKS}}$ and the \textit{maximum likelihood estimator} (MLE) denoted ${}_0\kappa_{\text{MLE}}$,
which is defined by maximization of the likelihood (\textit{i.e.} an implicit assumption of a flat prior on $\kappa$). Suppose the
noise properties are assumed to be Gaussian (as is common),
then the likelihood is given by
\begin{equation}
P(\gamma | \kappa) \propto | \Sigma |^{-1/2} \exp^{-\frac{1}{2} \chi^2} ,
\end{equation}
for $\chi^2 \equiv (\gamma - \bm{\Phi} \kappa)^T \Sigma^{-1} (\gamma - \bm{\Phi} \kappa)$ where
$\bm{\Phi} $ is simply the forward model and $\Sigma$ is the noise covariance.
The solution which minimizes the likelihood is thus given by
\begin{equation}
 {}_0\kappa_{\text{MLE}} = (\bm{\Phi} ^T \Sigma^{-1} \bm{\Phi} )^{-1} \bm{\Phi} ^T \Sigma^{-1} \gamma.
\end{equation}

Therefore for the idealised SKS estimator $\bm{\Phi} ^{-1}\gamma$ to be equivalent to the MLE estimator we require
$\bm{\Phi} ^{-1}$  and $(\bm{\Phi} ^T \Sigma^{-1} \bm{\Phi} )^{-1} \bm{\Phi} ^T \Sigma^{-1}$ to be equivalent operators.
Provided $\Phi$ is invertible the $\Sigma$ terms above trivially cancel resulting in the remaining
terms $(\bm{\Phi} ^T \bm{\Phi} )^{-1} \bm{\Phi} ^T$, which reduce to the idealised SKS estimator -- note that in the
idealised setting $\Phi$ is straightforwardly invertible given both the spherical harmonic transform and equation \ref{eq:spherical_kernel} are
invertible (ignoring the monopole $\ell=0$). Thus in this setting the idealised SKS and MLE estimator are equivalent.

However, in practical applications the forward model (\textit{e.g.} with PSF corrections, complex masking, \textit{etc.})
is unlikely to be invertible and hence the SKS and MLE estimators differ in practice.
Furthermore, due to limited observation quality (discussed in section \ref{sec:Noise_computation})
the noise covariance is typically large in magnitude. In such settings the inverse problem is
strongly ill-posed and thus significant regularization (introduced through the prior term) is
required to stabilize the inversion.
As such, a flat prior (MLE) results in unregularizied solutions which are highly  unlikely to perform well (noise
present in $\gamma$ is very likely to propogate directly into the $\kappa$ estimate). This noise
propogation is well-known, hence the SKS estimator used in practice always includes a smoothing
post-processing step (convolution with an arbitrary smoothing kernel) in an attempt to mitigate this noise.
Consequently, the SKS estimator used in practice does not support a principled statistical interpretation.

\section{Spherical Bayesian Mass-mapping} \label{sec:bayesian_inference}
Hierarchical Bayesian frameworks facilitate a natural, mathematically principled approach
to uncertainty quantification. For an elegant and approachable introduction to Bayesian
methods see \citet{[51]}. This section introduces Bayesian inference and proceeds to
demonstrate how one may cast the spherical mass-mapping inversion as a hierarchical
Bayesian inference problem. For notational ease, we drop spin subscripts on $\kappa$
and $\gamma$ henceforth.

\subsection{Bayesian Inference}

First consider the \textit{posterior distribution} given by Bayes' Theorem,
\begin{equation} \label{eq:bayes}
p(\kappa|\gamma; \mathcal{M}) = \frac{p(\gamma|\kappa; \mathcal{M})p(\kappa; \mathcal{M})}{\int_{\mathbb{C}^N} p(\gamma|\kappa; \mathcal{M})p(\kappa; \mathcal{M})d\kappa},
\end{equation}
where the \textit{likelihood function} $p(\gamma|\kappa; \mathcal{M})$ represents the
probability of observing a shear field $\gamma$ given a convergence field $\kappa$ and
some well defined model $\mathcal{M}$ (which includes both the mapping
$\bm{\Phi} : \kappa \mapsto \gamma$ and some assumptions of the noise model). The
second term in the numerator, $p(\kappa; \mathcal{M})$ is referred to as the
\textit{prior} which encodes some \textit{a priori} knowledge as to the nature of
$\kappa$. Finally, the integral denominator is the \textit{Bayesian evidence} (or
\textit{marginal likelihood}) which can be used for model comparison, though we do not
consider this within the scope of the current paper.
\par
One approach to estimate the convergence field is given by maximizing the posterior
odds conditional on the measurements $\gamma$ and model $\mathcal{M}$. Such a
solution is referred to as the \textit{maximum a posteriori} (MAP) solution,
$\kappa^{\text{map}}$. This can done by either maximization of the posterior or \emdash
due to the monotonicity of the logarithm function \emdash minimization of the log-posterior,
\begin{equation} \label{eq:log_posterior}
\argmaxT_{\kappa} \big \lbrace p(\kappa|\gamma;\mathcal{M}) \big \rbrace \equiv \argminT_{\kappa} \big \lbrace -\log ( \; p(\kappa|\gamma;\mathcal{M}) \;) \big \rbrace.
\end{equation}
This is a particularly helpful realization as the latter problem is more straightforward to
compute and, for log-concave posteriors, allows one to to pose the problem as a convex
optimization problem for which one may draw on the field of convex optimization.

\subsection{Spherical Sparse Mass-Mapping} \label{sec:forward_model}
In this paper we consider the ill-posed linear inverse problem of recovering the complex
discretized spherical convergence $\kappa \in \mathbb{C}^{N_{\mathbb{S}^2}}$ on the
complex $\mathbb{S}^2$-sphere from a typically incomplete ($M < N$) set of $M$ complex
discretized shear measurements $\gamma \in \mathbb{C}^M$. Throughout we adopt the
McEwen-Wiaux (MW) pixelization scheme, which provides theoretically exact spin spherical
harmonic transforms (SSHT) due to exact quadrature \citep{[52]}.
\par
We begin by defining the \textit{measurement operator} (operator which encodes the forward
model) which maps from a fiducial convergence field to the observed shear measurements
\begin{equation}
\bm{\Phi} \in \mathbb{C}^{M \times N_{\mathbb{S}^2}} : \kappa \in \mathbb{C}^{N_{\mathbb{S}^2}} \mapsto \gamma \in \mathbb{C}^M.
\end{equation}
In the spherical setting, by noting the spherical lensing forward model given by equation
(\ref{eq:spherical_kernel}) this measurement operator naturally takes the form,
\begin{equation} \label{eq:measurement_operator}
\bm{\Phi} = \bm{M} {}_2\tilde{\bm{Y}} \; \bm{W} \; {}_0 \bm{Y},
\end{equation}
where ${}_s\bm{Y}$ and ${}_s\tilde{\bm{Y}}$ represent the forward and inverse spin-$s$
spherical harmonic transforms respectively, $\bm{M}$ is a masking operator, and $\bm{W}$ is
harmonic space multiplication by the kernel $\mathcal{W}_{\ell}$ defined in equation
(\ref{eq:spherical_kernel}). The adjoint-measurement operator can then be shown
to be,
\begin{equation}
\bm{\Phi}^{\dagger} = \; {}_0 \bm{Y}^{\dagger} \; \bm{W} \; {}_2\tilde{\bm{Y}}^{\dagger}
\bm{M}^{\dagger},
\end{equation}
where it should be noted that from symmetry $\bm{W}$ is trivially self-adjoint. Additionally,
it is important to note that adjoint ($\dagger$) spin-s spherical harmonic transforms are not
equivalent to the corresponding inverse spherical harmonic transforms \emdash an important
caveat often overlooked throughout the field.

\subsubsection{Likelihood Function} \label{sec:Likelihood_function}

Suppose now that measurements $\gamma$ are acquired under some additive Gaussian
noise $n_i \sim \mathcal{N}(0,\sigma_i^2) \in \mathbb{C}^{M}$ where $\sigma_i$ is the
noise standard deviation of a given pixel which is primarily dependent on the number of
observations within said pixel, which is in turn dependent on the pixel size and
number density of galaxy observations. Then the data acquisition model is simply given by
\begin{equation}
\gamma = \bm{\Phi} \kappa + n.
\end{equation}
In such a setting the Bayesian likelihood function (data fidelity term) is given by the
product of Gaussian likelihoods defined on each pixel with pixel noise variance
$\sigma_i^2$, which is to say an overall multivariate Gaussian likelihood of known
covariance $\Sigma = \text{diag}(\sigma_1, \sigma_2,\dots,\sigma_M) \in \mathbb{R}^{M \times M}$.
Let $\Phi_i\kappa$ be the value of $\bm{\Phi}\kappa$ at pixel $i$, then the overall
likelihood is then defined as,
\begin{align}
p(\gamma|\kappa) &\propto \prod_{i=0}^{M} \exp \Bigg(\frac{-( \Phi_i \kappa - \gamma_i )^2}{2\sigma_i^2} \Bigg)
= \prod_{i=0}^{M} \exp \Bigg( \frac{-1}{2} \Big ( \bar{\Phi}_i \kappa - \bar{\gamma}_i \Big )^2 \Bigg), \nonumber \\
&= p(\gamma | \kappa) \propto \exp \Bigg( \frac{-\norm{ \bar{\bm{\Phi}} \kappa - \bar{\gamma} }_2^2}{2} \Bigg), \label{eq:covariance_likelihood}
\end{align}
where $\norm{ \cdot}_2$ is the $\ell_2$-norm and $\bar{\bm{\Phi}} = \Sigma^{-\frac{1}{2}}\bm{\Phi}$
is a composition of the measurement operator and an inverse covariance weighting as
defined in Section \ref{sec:forward_model}. Effectively this covariance weighting leads to
measurements $\bar{\gamma} = \Sigma^{-\frac{1}{2}}\gamma$ which whiten the typically
non-uniform noise variance in the observational data (shear field).
\par
This likelihood is therefore structured to correctly account for the covariance of
observational data. In this case the covariance matrix is taken to be diagonal but not
necessarily proportional to the identity matrix \emdash therefore accounting for varied
numbers of observations per pixel. There are several points which should be noted. In the
above we have explicitly ignored the complicating factor of intrinsic galaxy alignments
which in practice would lead to non-diagonal covariance. This extension can easily be
supported, given a sound understanding of the effects of intrinsic alignments on the
data covariance (which in practice may be challenging).
\par
Additionally here we, for simplicity, assume each pixel contains a sufficient number of galaxy
observations that a \textit{central limit theorem} (CLT) argument for pixel noise can be justified.
Largely this assumption is acceptable, however as the resolution
increases (pixel size decreases) the noise becomes increasingly non-Gaussian.
\par
Finally, the forward model considered here (Section \ref{sec:forward_model}) begins from
$\kappa$ and ends at masked, gridded $\gamma$ measurements, however there are
several steps which must take place before one acquires such measurements. One may
therefore wish to extend this model to incorporate such complicating factors as pixelisation
effects, reduced shear (see section \ref{sec:reduced_shear}), point squared function (PSF) errors \textit{etc.}
\par
It should then be explicitly noted that this mass-mapping formalism requires only that the
posterior belong to the (rather comprehensive) set of log-concave functions, and as such
one can directly interchange the noise model or introduce complicating factors where
desired provided the posterior remains log-concave.

\subsubsection{Prior Function} \label{sec:Prior_function}

As this inverse problem is ill-posed (often seriously), maximum likelihood estimators (MLE)
are sub-optimal and must be regularized by some prior assumption as to the nature of the
convergence field. In this work we select a sparsity promoting, Laplace-type prior in the
form of the $\ell_1$-norm $\norm{.}_1$ \emdash though as discussed in section
\ref{sec:Likelihood_function} this formalism supports any log-concave priors of which there
are many to choose from (\textit{e.g.} most exponential family priors).
\par
Laplace-type priors are often adopted when one wishes to promote sparsity in a given
dictionary or basis. Wavelets $\bm{\Psi}$ are localised in both the
frequency and spatial domains and thus constitute a naturally sparsifying dictionary for most
physical signals. There are several wavelet constructions on the sphere that may be considered
\citep[see \textit{e.g.}][]{schroder:1995, barreiro:2000, starck2006wavelets, mcewen:2008:fsi, mcewen:szip, wiaux2008, McEwenS2DWLocalisation, narcowich2006localized,
baldi2009asymptotics, marinucci2007, mcewen:s2let_ridgelets, Chan2017Curvelets} with varying localisation and uncorrelation properties.
In this paper we adopt the scale-discretised wavelets \citep{wiaux2008,Leistedt2013, S2DW_2013, [48]} scheme as not only does it satisfy qausi-exponential locatlisation and asymptotic uncorrelation properties \citep{McEwenS2DWLocalisation} but also supports directionality which may often be
of interest for the weak lensing setting.

We specifically adopt a Laplace-type wavelet log-prior
$\norm{\bm{\Psi}(\cdot)}_1$.
Note that as $\norm{\cdot}_1$ is a discretization of the continuous $\ell_1$-norm it must
be reweighted by wavelet pixel size, which in practice is as simple as multiplying a given
wavelet coefficient by a factor proportional to $\sin(\theta)$ where $\theta$ is the angular
deviation  of the given pixel from the pole. Throughout this paper any reference to the
$\ell_1$-norm applied to a spherical space refers explicitly to this spherically reweighted norm.
\par
With our choice of $\ell_1$-norm regularization the prior can be written compactly as
\begin{equation}
  p(\kappa) \propto \exp \Big(-\mu \norm{\tilde{\bm{\Psi}}^{\dag}\kappa}_1 \Big), \label{eq:prior}
\end{equation}
where $\tilde{\bm{\Psi}}^{\dag}$ is the analysis forward-adjoint spherical wavelet transforms
(see equation \ref{eq:wavelet_transforms} in the appendix) with coefficients $\tilde{\Psi}^{\dag}_i$, and
$\mu \in \mathbb{R}^+$ is the regularization parameter. It is assumed here that the spherical
wavelet dictionary $\tilde{\bm{\Psi}}$ is a naturally sparsifying dictionary for the convergence
field defined on the sphere. In practice one may select whichever dictionary one's prior
knowledge of the convergence indicates is likely to be highly sparsifying.
\par
Conceptually, a sparsity-promoting prior can be though of as a mathematical manifestation
of \textit{Occam's Razor} \emdash the philosophical notion that the simplest answer is
usually the best answer. Mathematically, this is equivalent to down-weighting solutions with
large numbers of non-zero coefficients, which may match the noisy data perfectly, in favour
of a less perfect match but with significantly fewer non-zero coefficients.
\par
Alternatively, one may view sparsity priors (in this context) as a relative assumption of the
sparsity of the true signal and noise signal when projected into a sparsifying dictionary. This is to say
that the assumption is that the noise signal will be less sparse in $\tilde{\bm{\Psi}}$ than the true signal.
Typically noise signals are relatively uniformly distributed in wavelet space, whereas most physical
signals are sparsely distributed and therefore this relative interpretation of the sparsity prior makes
reasonable sense \citep{Mallat_textbook}.
\par
Note that the only constraint on the posterior is that it must be log-concave (such that the
log-posterior is convex). Hence one can select any log-concave prior within this
framework, \textit{e.g.} one could select an $\ell_2$-norm prior which with minor
adjustments produces Wiener filtering \citep[see][for alternate iterative Wiener filtering
approaches]{Horowitz2018}, or a flat prior which produces the
\textit{maximum likelihood estimate} (MLE).

\subsection{Implementation} \label{sec:numerical_implementation}

The minimization of the log-posterior in equation (\ref{eq:log_posterior}) is (in the analysis
setting) therefore precisely the same as solving,
\begin{equation} \label{eq:optimization}
\kappa^{\text{map}} = \argminT_{\kappa} \underbrace{  \Bigg \lbrace \mu \norm{ \bm{\Psi}^{\dag}\kappa}_1 + \frac{\norm{\bar{\bm{\Phi}} \kappa - \bar{\gamma}}_2^2}{2} \Bigg \rbrace.}_{\text{Objective function}}
\end{equation}
The bracketed term on the RHS is referred to as the \textit{objective function}.
We solve this convex optimization problem using the S2INV \citep{S2INV} code which is largely built
around the SOPT C++ object oriented framework\footnote{https://github.com/astro-informatics/sopt} \citep{SOPT, Onose_2016_SOPT, [39], Carrillo_2013_SOPT},
utilizing an adapted proximal forward-backward splitting algorithm \citep{[53]}, although a
variety of alternate algorithms are provided within S2INV.  Wavelet transforms on the sphere are computed using S2LET\footnote{http://astro-informatics.github.io/s2let/} \citep{[52], [54], Leistedt2013, [42], [48],  McEwenS2DWLocalisation}, which in turn makes use of
SSHT\footnote{https://astro-informatics.github.io/ssht/} \citep{[52], [41]} to compute spherical harmonic transforms and SO3\footnote{http://astro-informatics.github.io/so3/} \citep{[48]} to compute Wigner transforms.

\par
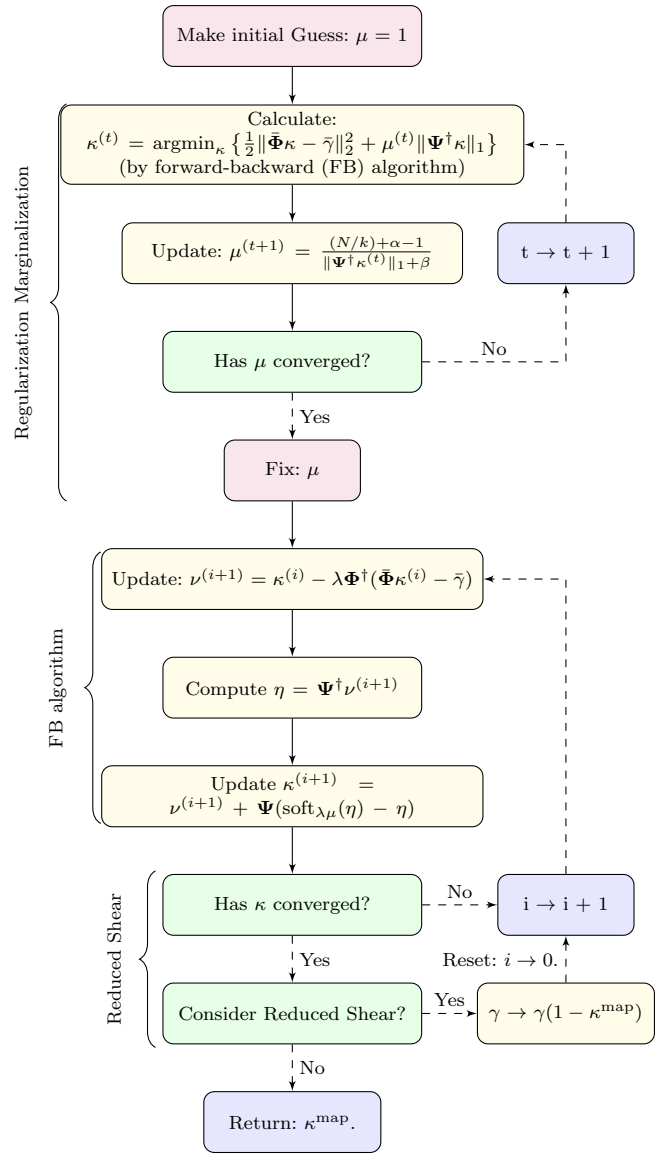
\begin{figure}
\begin{center}
\begin{tikzpicture}[node distance = 2cm, auto, scale=0.9, every node/.style={scale=0.9}]
    \node [parameter_fixed, text width=12em] (1) {Make initial Guess: $\mu = 1$};
    \node [block, below of=1, node distance=1.6cm, text width=22em] (2) {Calculate: \\
    $\kappa^{(t)} = \argminT_{\kappa}  \big \lbrace \frac{1}{2}\norm{\bar{\bm{\Phi}} \kappa - \bar{\gamma}}_2^2 + \mu^{(t)} \norm{ \bm{\Psi}^{\dag}\kappa}_1 \big \rbrace$
    \\
    (by forward-backward (FB) algorithm)};
    \node [block, below of=2, node distance=1.6cm, text width=16em] (3) {Update: $\mu^{(t+1)} = \frac{(N/k)+\alpha-1}{\norm{ \bm{\Psi}^{\dag}\kappa^{(t)}}_1 + \beta}$};
    \node [decision, below of=3, node distance=1.6cm, text width=12em] (4) {Has $\mu$ converged?};
    \node [iteration, right of=3, node distance=4.0cm, text width=6em] (5) {t $\rightarrow$ t + 1};
    \node [parameter_fixed, below of=4, node distance=1.6cm, text width=6em] (6) {Fix: $\mu$};
    \node [block, below of=6, node distance=1.6cm, text width=18em] (7) {Update: $\nu^{(i+1)} = \kappa^{(i)} - \lambda \bm{\Phi}^{\dagger}(\bar{\bm{\Phi}} \kappa^{(i)} - \bar{\gamma})$};
    \node [block, below of=7, node distance=1.6cm, text width=12em] (8a) {Compute $\eta = \bm{\Psi}^{\dagger}\nu^{(i+1)}$};
    \node [block, below of=8a, node distance=1.6cm, text width=18em] (8b) {Update $\kappa^{(i+1)} = \nu^{(i+1)} + \bm{\Psi} ( \text{soft}_{\lambda\mu}(\eta)-\eta)$};
    \node [decision, below of=8b, node distance=1.6cm, text width=12em] (9) {Has $\kappa$ converged?};
    \node [iteration, right of=9, node distance=4.0cm, text width=6em] (10) {i $\rightarrow$ i + 1};
    \node [decision, below of=9, node distance=1.6cm, text width=12em] (11) {Consider Reduced Shear?};
    \node [block, right of=11, node distance=4.0cm, text width=8em] (12) {$\gamma \rightarrow \gamma ( 1 - \kappa^{\text{map}})$};
    \node [iteration, below of= 11, node distance=1.6cm, text width=8em] (13) {Return: $\kappa^{\text{map}}$.};

    \draw [decorate,decoration={brace,amplitude=6pt},xshift=0pt,yshift=0pt]
    (-2.0,-14.9) -- (-2.0,-12.3) node [black,midway,xshift=-0.4cm]
    {\footnotesize \rotatebox{90}{Reduced Shear}};

    \draw [decorate,decoration={brace,amplitude=6pt},xshift=0pt,yshift=0pt]
    (-2.8,-11.6) -- (-2.8,-7.6) node [black,midway,xshift=-0.4cm]
    {\footnotesize \rotatebox{90}{FB algorithm}};

    \draw [decorate,decoration={brace,amplitude=6pt},xshift=0pt,yshift=0pt]
    (-3.3,-6.85) -- (-3.3,-1.0) node [black,midway,xshift=-0.4cm]
    {\footnotesize \rotatebox{90}{Regularization Marginalization}};

    \path [line] (1) -- (2);
    \path [line] (2) -- (3);
    \path [line] (3) -- (4);
    \path [line,dashed] (4) -| node[near start] {No}(5);
    \path [line,dashed] (5) |- (2);
    \path [line,dashed] (4) -- node {Yes}(6);
    \path [line] (6) -- (7);
    \path [line] (7) -- (8a);
    \path [line] (8a) -- (8b);
    \path [line] (8b) -- (9);
    \path [line,dashed] (9) -- node {No}(10);
    \path [line,dashed] (10) |- (7);
    \path [line,dashed] (9) -- node {Yes} (11);
    \path [line,dashed] (11) -- node {Yes} (12);
    \path [line,dashed] (11) -- node {No} (13);
    \path [line,dashed] (12) -- node [left]{Reset: $i \rightarrow 0$.}(10);
\end{tikzpicture}
\caption{Schematic of proximal forward-backward splitting algorithm used \citep{[53]}.
Note that the first iterative block represents the Majorize-Minimization (MM) algorithm marginalization over
the regularization parameter (which here is treated as a nuisance parameter), the
second itertaive block represents the primary proximal forward backward iterations,
and the final (optional) block represents the reduced shear outer iterations.
Note that the $\text{soft}_{\lambda, \mu}(\eta)$ operation is the soft thresholding operation,
which is the proximal projection of the $\ell_1$-norm \citep[see \textit{e.g.} ][ for details]{[10], [11], [12]}.
}
\label{fig:forward-backwards-splitting}
\end{center}
\end{figure}

To deal with the non-differentiable $\ell_1$-norm prior, gradient operators $\nabla$ are in some sense
replaced by proximal operators when applied to the non-differentiable term \citep{Moreau1962}.
The iteration steps are provided in the schematic of Figure \ref{fig:forward-backwards-splitting}, for full
details of the derivation of the proximal forward-backward algorithm iterations look to \cite{[53]}.
These primary optimizations are terminated once the objective function is updated by less
than a set threshold (in our experiments $10^{-6})$ between iterations.

\subsubsection{Reduced shear} \label{sec:reduced_shear}
Figure \ref{fig:forward-backwards-splitting} displays a schematic representation of the
steps taken in computing $\kappa^{\text{map}}$. A degeneracy between the
convergence field $\kappa$ and shear field $\gamma$ exists, and as such $\gamma$ is
not a true observable. Instead the \textit{reduced shear} $g$ is the true observable where,
\begin{equation}
g(\omega) = \frac{\mathbf{\gamma}(\omega)}{1 - \mathbf{\kappa}(\omega)}.
\end{equation}
When working sufficiently within the weak lensing regime $\kappa \ll 1$ and
$\gamma \approx g \ll 1$. Although typically the reduced shear need not be accounted for,
for completeness we correct for the reduced shear \citep{Mediavilla_2016_book, [3],M1}.
We add correcting iterations outside our primary iterations to maintain the linearity of the
overall reconstruction. Our reduced shear correction iterations are displayed schematically
in the final loop of Figure \ref{fig:forward-backwards-splitting}.

Reduced shear
iterations are deemed to have converged once the convergence update $\maxT_j |\kappa_j^{(i)} -\kappa_j^{(i+1)}|  < 10^{-10} $
where $j$ runs over all pixels \citep[as in][]{[3]}.

\subsubsection{Bayesian Regularization Parameter} \label{sec:standard_MAP_regularisation}
For recovered statistics to be truly principled, the regularization parameter must necessarily
be folded into the hierarchy or correctly marginalized over. One way to do this was recently
developed \citep{[16]} and shown to work well in the planar weak lensing setting \citep{M1}.
\par
This Bayesian hierarchical inference approach assumes a gamma distribution hyper-prior
\begin{equation}
p(\mu) = \frac{\beta^{\alpha}}{\Gamma(\alpha)}\mu^{\alpha - 1} e^{-\beta \mu} \mathbb{I}_{\mathbb{R}^+}(\mu),
\end{equation}
with weakly dependent hyper-parameters $\alpha$ and $\beta$ which without loss of generality
can be fixed at $\alpha = \beta = 1$. We then iterate \citep{[16]} to effectively
marginalize over $\mu$ which is treated as a nuisance parameter in the main body of our hierarchy.
These iterations are,
\begin{equation}
\kappa^{(t)} = \argminT_{\kappa}  \big \lbrace \frac{1}{2}\norm{\bar{\bm{\Phi}} \kappa -
\bar{\gamma}}_2^2 + \mu^{(t)} \norm{ \bm{\Psi}^{\dag}\kappa}_1 \big \rbrace,
\end{equation}
\begin{equation} \label{eq:MAP_regularization_term}
\mu^{(t+1)} = \frac{(N/k)+\alpha-1}{\norm{ \bm{\Psi}^{\dag}\kappa^{(t)}}_1 + \beta}
\end{equation}
where the log-prior $\norm{ \bm{\Psi}^{\dag}\kappa}_1$ is $k$-homogeneous. Note that a
sufficient statistic (log-prior) is k-homogeneous if $\exists \; k \in \mathbb{R}^+$ such
that,
\begin{equation}
f(\eta x) = \eta^kf(x), \: \forall x \in \mathbb{R}^M, \: \forall \eta > 0.
\end{equation}
Further note that all norms, composite norms, and compositions of norms with linear operators are
1-homogeneous, \textit{i.e.} $k=1$. See \cite{[16]} for further details. These regularization marginalization
iterations are terminated when the update to the regularization parameter is less than 1\%
\textit{i.e.} $|\mu^{(i+1)} - \mu^{(i)}| / \mu^{(i)} < 0.01$.

\subsubsection{Computational efficiency}
As discussed in \ref{sec:numerical_implementation} all iterations consist of a forward step
which includes application of the measurement operator before computing the data fidelity
term, followed by the backward step which includes application of the spherical wavelet
transform.
\par
The measurement operator is dominated by the spin spherical harmonic transforms which
scale as $\mathcal{O}(L^3)$. Similarly the computational efficiency of the wavelet transform
is dominanted by underlying harmonic transforms, however with directionality
$N$ (\textit{i.e.} wavelet on the rotation group) the transform scales as
$\mathcal{O}(N \times L^3)$. The overall forward-backward algorithm scales additively as
$\mathcal{O}(K \times (N+1) \times L^3) \sim \mathcal{O}(K \times N \times L^3)$ where $K$
is the total number of iterations required for convergence.
\par
The SKS operator also requires the application of spin spherical harmonic transforms and
therefore scales as $\mathcal{O}(L^3)$. However the SKS method requires only a single
application of the transform and thus the ratio of computational efficiency between the two
algorithms effectively scales as $\mathcal{O}(K \times N)$ \emdash which is to say the
difference in computational efficiency is primarily determined by the choice of wavelet complexity
and the magnitudes of the associated convergence criteria.
\par
In practice, including the marginalization preliminary iterations and subsequent annealing
iterations to optimize convergence, we find $\mathcal{O}(10^2)$ iterations are sufficient for
convergence. We consider axisymmetric wavelets ($N=1$) in this article, thus the DarkMapper
algorithm is $\mathcal{O}(10^2)$ times slower than SKS but with greatly superior reconstruction
performance and the ability to quantify uncertainties in a statistically principled manner.
\par
It is interesting to note that MCMC methods typically require a very large number of
samples, with each individual sample requiring at least one spin spherical harmonic
transform. Therefore the increase in computational efficiency of this approximate Bayesian
inference over sampling methods is roughly given by $\mathcal{O}(n_\text{samples} / 10^2)$
where $n_\text{samples}$ is the total number of samples required for convergence of a
given MCMC sampling method. As MCMC methods often require at least
$\mathcal{O}(10^6)$  this increase in computation speed is (many) orders of magnitude. In
the spherical setting an $\mathcal{O}(10^4)$ increase in computation speed results in
computations which would take $\mathcal{O}$(decades) taking $\mathcal{O}$(days).

\section{Bayesian Uncertainty Quantification} \label{sec:UQ}
Though MAP estimates provide high fidelity estimates of the convergence field uncertainties
on these estimates are a necessity if one aims to make statistically principled inferences.
Generally, for scientific inference one should prioritize principled uncertainties over image
aesthetics.
\par
Bayesian inference approaches as presented in Section \ref{sec:bayesian_inference}
provide principled statistical frameworks through which quantification of uncertainty on
recovered statistics comes naturally from the posterior. Typically the posterior cannot be
evaluated analytically, and so \textit{Markov Chain Monte Carlo} (MCMC) sampling
methods must be used. In moderate to low dimensional settings for computationally cheap
functions, MCMC chains can feasibly be computed. However, in high dimensional spherical
settings, MCMC techniques quickly become challenging to compute.
\par
Bespoke MCMC techniques have been developed for the weak lensing setting
\citep[\textit{e.g.}][]{[56],[57],[58]} which can improve computational efficiency, yet these
methods will find it challenging to accommodate future `Big Data' from high-resolution
wide-field surveys. Furthermore, these sometimes come with additional restrictions (\textit{e.g.}
some are restricted to Gaussian priors). This provides strong motivation for the development
of fast, approximate Bayesian inference approachs \cite{[10],[12],M1,M2,M3},
the uncertainty quantification of which we extend to the complex $\mathbb{S}^2$-sphere
and present in this section.

\subsection{Highest Posterior Density Region}
At $100(1-\alpha)\%$ confidence a sub-set $C_{\alpha} \in \mathbb{C}^{N_{\mathbb{S}^2}}$
of the posterior space is considered a credible region of the posterior \textit{iif} the integral equation,
\begin{equation} \label{eq:Credible_Integral}
p(\kappa \in C_{\alpha}|\gamma) = \int_{\kappa \in \mathbb{C}^{N_{\mathbb{S}^2}}} p(\kappa|\gamma)\mathbb{I}_{C_{\alpha}}d\kappa = 1 - \alpha,
\end{equation}
is satisfied, where we have used the set indicator function $\mathbb{I}_{C_{\alpha}}$,
defined to be,
\begin{equation} \label{eq:indicator}
  \mathbb{I}_{C_{\alpha}}=\begin{cases}
               1 \quad \text{if,}\quad  \kappa \in C_{\alpha} \\
               0 \quad \text{if,}\quad \kappa \not\in C_{\alpha}.\\
            \end{cases}
\end{equation}
Theoretically there are infinitely many regions which satisfy the integral in equation
(\ref{eq:Credible_Integral}). However, the decision-theoretic optimal region \emdash in the
sense of minimum volume \emdash is the \textit{Highest Posterior Density} (HPD)
credible-region, which is given by \citep{[19]}
\begin{equation}
C_{\alpha} := \lbrace \kappa : f(\kappa) + g(\kappa) \leq \epsilon_{\alpha} \rbrace,
\end{equation}
where the combination $f(\kappa) + g(\kappa)$ is our objective function derived in
Section \ref{sec:numerical_implementation}, and $\epsilon_{\alpha}$ is an isocontour
(\textit{i.e.} level-set) of the log-posterior.
\par
However, in high dimensional $(N \gg 1)$ settings $\epsilon_\alpha$ (and therefore
$C_\alpha$) becomes particularly difficult to compute, thus motivating the development of
alternate approaches that are fast and approximate. Recent advances in probability concentration
theory have led to the derivation \citep{[10]} of a conservative approximate credible-region
$C_{\alpha}^{\prime}$ for log-concave distributions. This approximate region is defined as,
\begin{equation} \label{eq:credible_set}
C^{\prime}_{\alpha} := \lbrace \kappa : f(\kappa) + g(\kappa) \leq \epsilon^{\prime}_{\alpha} \rbrace,
\end{equation}
such that,
\begin{equation} \label{eq:level_set_threshold}
\epsilon^{\prime}_{\alpha} = f(\kappa^{\text{map}}) + g(\kappa^{\text{map}}) + \tau_{\alpha} \sqrt{N} + N,
\end{equation}
is the approximate level-set threshold with constant $\tau_{\alpha} = \sqrt{16 \log(3 / \alpha)}$.
Recall that $N$ is the dimension of $\kappa \in \mathbb{C}^{N_{\mathbb{S}^2}}$ which for
equiangular spherical sampling \citep[MW; ][]{[52]} is given by,
\begin{equation}
N_{\text{MW}} \equiv  \ell_{\text{max}} ( 2 \ell_{\text{max}} - 1 ) \approx 2 \ell_{\text{max}}^2,
\end{equation}
for signals with angular band-limit $\ell_{\text{max}}$. An upper bound on the error
introduced through this approximation has been shown to exist \citep{[10]} and is given by,
\begin{equation}
0 \leq \epsilon^{\prime}_{\alpha} - \epsilon_{\alpha} \leq \eta_{\alpha} \sqrt{N} + N,
\end{equation}
where $\eta_{\alpha} = \sqrt{16 \log (3/\alpha)} + \sqrt{1/\alpha}$. This error scales at most
linearly with $N$ and in high dimensional settings can be somewhat large, though in
practice we find this error upper-bound to be extremely conservative \citep{M2}.
\par
Note that the error is positive semi-definite which corroborates the assertion that
$C_{\alpha}^{\prime}$ is a conservative approximation. Mathematically, this is to say that
the true HPD credible region $C_{\alpha}$ is sub-set of the approximate HPD credible
region $C_{\alpha}^{\prime}$ \textit{i.e.} $C_{\alpha} \subseteq C_{\alpha}^{\prime}$. This
ensures that if some convergence field $\kappa \not\in C_{\alpha}^{\prime}$ then necessarily
$\kappa \not\in C_{\alpha}$.
\par
Further note that although we adopt the approximate level-set threshold derived in
\citet{[10]} in this work, research into these types of bounds is a relatively new area of study.
Thus, if and when new (more constraining) bounds are derived they can trivially be
substituted here.

\subsection{General Application}
Having introduced the concept of an approximate HPD-credible region $C^{\prime}_{\alpha}$
of the posterior, the question then immediately arises as to how one can utilize this
information in practice. In MCMC sampling type approaches, one may simply use the
recovered samples to quantify uncertainty at well defined confidence on specific properties
of the recovered posterior. In our setting, we have recovered only the MAP solution in a form
which supports trivial computation of the approximate level-set threshold, and thus
$C^{\prime}_{\alpha}$.
\par
With such limited posterior knowledge one may only ask whether a surrogate solution
(an adjusted convergence map) does or does not belong to $C^{\prime}_{\alpha}$. In effect
this is to say that in our formalism all questions of the posterior must be cast as Bayesian
hypothesis tests of varying complexity \citep{M1, M3}. Some examples are provided in
the following subsections.

\subsubsection{Bayesian hypothesis testing} \label{sec:Hypothesis_testing}
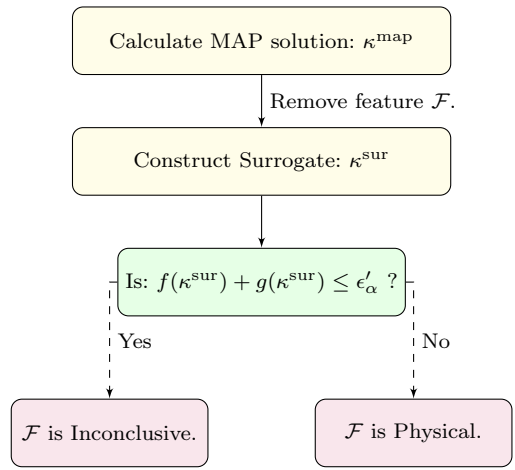
\begin{figure}
\begin{center}
\begin{tikzpicture}[node distance = 2cm, auto]
    \node [block, text width=16em] (1) {Calculate MAP solution: $\kappa^{\text{map}}$};
    \node [block, below of=1, node distance=1.6cm, text width=16em] (2) {Construct Surrogate: $\kappa^{\text{sur}}$};
    \node [decision, below of=2, node distance=1.6cm] (3) {Is: $f(\kappa^{\text{sur}}) + g(\kappa^{\text{sur}}) \leq \epsilon_{\alpha}^{\prime}$ ? };
    \node [parameter_fixed, below of=3, right of=3, text width=8em] (4) {$\mathcal{F}$ is Physical.};
    \node [parameter_fixed, below of=3, left of=3, text width=8em] (5) {$\mathcal{F}$ is Inconclusive.};

    \path [line] (1) -- node {Remove feature $\mathcal{F}$.} (2);
    \path [line] (2) -- (3);
    \path [line,dashed] (3) -| node [near end] {No}(4);
    \path [line,dashed] (3) -| node [near end] {Yes}(5);

\end{tikzpicture}
\caption{Schematic of hypothesis testing \citep{M1}. The feature $\mathcal{F}$ is entirely
general and can be constructed by any well defined operator on the MAP solution
$\kappa^{\text{map}}$. }
\label{fig:HypthosisTesting}
\end{center}
\end{figure}

A Bayesian hypothesis test on the posterior (see Figure \ref{fig:HypthosisTesting}) is simply:
the MAP convergence is recovered, a feature of that map is removed\footnote{In practice
one may simply adjust $\kappa^{\text{sur}}$ to suit a specific question of the posterior.} to
form a surrogate map $\kappa^{\text{sur}}$, if $\kappa^{\text{sur}} \not\in C^{\prime}_{\alpha}$,
then $\kappa^{\text{sur}} \not\in C_{\alpha}$, and thus the hypothesis that feature of interest is
insignificant is rejected at some well defined confidence, implying that the feature cannot be
deemed insignificant at said confidence \citep[for more details look to ][]{[12], M1}.
\par
One can invisage constructing substantially more complicated uncertainty quantification
techniques \textit{via} iterative application of Bayesian hypothesis testing or by more
complicated individual Bayesian hypothesis tests.

\subsubsection{Local credible intervals}

The next most straightforward uncertainty quantification technique is given by the notion of
local credible intervals \citep{[12], M2} which are in effect pixel-level Bayesian uncertainty
(error) bars on recovered maps. Conceptually these are formed by splitting the recovered
MAP estimate into superpixels (groups of adjacent pixels), then within each superpixel
(keeping all other pixels fixed at their MAP values) iteratively increasing (decreasing) the
recovered pixel intensity and thus constructing surrogate solutions, checking whether these
surrogate solutions belong to $C^{\prime}_{\alpha}$. Once the maximum (minimum)
super-pixel intensity is located (typically \textit{via} bisection) the difference (maximum -
minimum) is taken to be the range of values which cannot be rejected for a recovered
super-pixels intensity \emdash hence the notion of these representing pixel level Bayesian
uncertainty (error) bars.

\subsubsection{Uncertainty quantification of global features}

For science, in particularly for cosmology, it is often perhaps more informative to leverage
the concept of Bayesian hypothesis testing to consider global structure, and therefore
consider global (or aggregate) statistics of a recovered field. To do so one must simply
define a logically consistant algorithm which constructs surrogate convergence solutions
that are representative of the global question they wish to ask of the recovered convergence
field, after which the process follows in much the same way as demonstrated for forming
local credible intervals.
\par
It should, however, be noted that one must be careful how one poses these global
questions, as the questions of interest are often inherently non-convex and must be
solved \textit{via} decision theory methods. A good example of how one can apply
hypothesis testing to global structure can be found in \citet{M3} where the Bayesian
uncertainty in the aggregate peak statitic is recovered.
\par
Here we have discussed only a few possible uncertainty quantification techniques which
are supported by this formalism, though in practice following the methodology outlined
above one can form uncertainty quantification techniques around a far more comprehensive
set of global features (or equally statistics) provided a few important caveats are understood:
the process of Bayesian hypothesis tests suggested to quantify a specified uncertainty are
well defined and clearly explained, the limitations of any method are fully acknowledged,
and the results are interpretted correctly so as to mitigate unjustified statistical statements.
We present a specific example on current cosmic shear data in Section \ref{sec:public_data}.

\subsubsection{The curse of dimensionality}

Finally it is academic to note that the concept of changing only a small number of pixels of
a given map whilst fixing the remaining at their MAP values is explicitly recovering
conditional probabilities which are by definition the largest possible uncertainties.
Though this is precisely what one requires of such approximations it highlights an inherent
drawback of such approaches. As the approximate level-set threshold scales
with the total dimension of the inference in high dimensional cases, the uncertainty of any
individual local structure within an image becomes large.

Conceptually this makes sense as the higher the dimensionality of the problem, the more
statistical fluctuations occur and thus the higher the chance that a
statistical fluctuation produced the feature of interest. As such, for anything higher than
moderate dimensional settings local uncertainties become very large and one
should prioritize global or aggregate statistics.

\section{Simulations} \label{sec:Sim_testing}
In this section we apply the spherical Kaiser-Squires (SKS) estimator , both with and without
post-processing smoothing, and the spherical sparse hierarchical Bayesian (DarkMapper)
estimator developed in this article to a range of realistic N-body simulations which are
masked throughout by a pseudo-Euclid mask so as to best match upcoming Stage
\rom{4} surveys.

\subsection{Data-set} \label{sec:dataset}
Throughout this article we perform reconstructions and uncertainty quantification on
simulated convergence maps generated from the high resolution Takahashi N-body
simulation datasets \citep{Ryuichi}\footnote{These datasets can be found at
\url{http://cosmo.phys.hirosaki-u.ac.jp/takahasi/allsky_raytracing/}.}. These mock
convergence maps are generated \textit{via} multiple-lens plane ray-tracing, and are
provided for a range of comoving distances. Specifically, simulated convergence maps
are presented at every $150$Mpc/h for redshift $z_s \in [0.05, 5.3]$. The cosmological
parameters selected for this suite of simulations are
$\Omega_m = 1, \Omega_{\Lambda} = 0.279, \Omega_{\text{cdm}} =  0.233, \Omega_b = 0.046, h=0.7, \sigma_8 = 0.82$
and $ns = 0.97$ which are consistent with the WMAP 9 year result \citep{WMAP_9year}.

We select redshift slice $16$ which corresponds to the slice with redshift $z_s \sim 1$.
To mitigate the Poisson noise present in such N-body snapshots we convolve the Takahashi
convergence with a very small smoothing kernel sufficient only to remove the noise whilst
adjusting the signal as little as possible. Finally we apply a pseudo-Euclid masking (a
straightforward masking of the galactic plane and the ecliptic) so as
to best mimic the setting of upcoming Stage \rom{4} surveys.

\begin{table*}
  	\centering
\begin{tabular}{|p{1cm}p{1cm}||p{1.3cm}p{1.3cm}||p{1.3cm}p{1.3cm}||p{1.3cm}p{1.3cm}||p{2.0cm}|}
\hline
\multicolumn{2}{|c||}{$\ell_{\text{max}} = 2048$}  & \multicolumn{2}{c||}{SKS} & \multicolumn{2}{c||}{SKS (optimal smoothing)} & \multicolumn{2}{c||}{DarkMapper} & \multicolumn{1}{c|}{Difference} \\
\hline
Setting & $n_{\text{gal}}$ & SNR (dB)   & P$_{\text{correlation}}$  & SNR (dB)  &  P$_{\text{correlation}}$ & SNR (dB)  &  P$_{\text{correlation}}$ & $\Delta$ SNR (dB) \\
\hline \hline
Stage \rom{3}  & 5 & -9.792  & 0.403  & 0.962 & 0.759 & 5.494  & 0.904 &  +15.286 (+4.532) \\
 & 10 & -6.794 & 0.532 & 1.108 & 0.806 & 7.299 & 0.935 & +14.093 (+6.191) \\
\hline
\rowcolor{LightCyan}
Stage \rom{4} & 30 & -2.091 & 0.732 & 1.254 & 0.854 & 9.767 & 0.964 & +11.858 (+8.513)  \\
\hline
Idealized  & 100 & 2.956 & 0.887 & n/a  & n/a &  12.132 & 0.980 &  +9.176 (n/a)\\
\hline
\end{tabular}
\caption{Numerical results from reconstructions of Takahashi simulations as discussed
in Sections \ref{sec:dataset} and \ref{sec:methodology}. In each case the DarkMapper
estimator drastically outperforms both the SKS estimator and the optimally smoothed SKS
estimator (which cannot in practice be achieved due to \textit{ad hoc} smoothing kernel selection)
in both recovered signal to noise ratio (SNR) and the Pearson correlation coefficient
P$_{\text{correlation}}$. Highlighted are the results most representative of the imminent
Stage \rom{4} surveys, such as Euclid and LSST. As Stage \rom{4} surveys forecast large
sky fractions to avoid projection effects \citep{[3], Zoe2019}
mass-mapping must be performed natively on the sphere. Thus this spherical mass-mapping
formalism is, at least currently, the optimal choice for Stage \rom{4} weak lensing mass-mapping.
Note that no post-processing by smoothing increased the recovered SNR for the idealized
$n_{\text{gal}} = 100$ setting for the SKS estimator and so was recorded as n/a.
}
\label{table:recon_fidelity}
\end{table*}%

\subsection{Methodology} \label{sec:methodology}
As in previous work \citep{M1,M2,M3} we begin by applying the measurement operator
$\bm{\Phi}$ (see equation \ref{eq:measurement_operator}) to the fiducial ground truth, full-sky Takahashi
convergence map $\kappa$ to create artificial masked clean shear measurements
$\gamma \in \mathbb{C}^M$.
\par
A noise standard deviation $\sigma_i$ is computed (see Section \ref{sec:Noise_computation})
for each pixel $i$ individually and used to construct known diagonal covariance
$\Sigma$.\footnote{Note we here do note consider off diagonal terms which may arise due
to intrinsic galaxy alignments though in future this can be incorporated straightforwardly.}
Hence we create noisy simulated shear observations $\gamma_n = \gamma + n$ and a
simulated data covariance $\Sigma$ which would in practice be provided by the observation
team \emdash this covariance is defined by the number of galaxy observations within a
given pixel of the sky.
\par
We then apply the standard SKS estimator and the DarkMapper estimator presented in this
paper to these noisy artificial measurements $\gamma_n$ to create estimates of the fiducial
convergence map $\kappa$. For DarkMapper we simply adopt diadic axisymmetric spherical
wavelets ($N=1$ and $\lambda=2$ for simplicity), with scale-discretised harmonic tiling \citep{McEwenS2DWLocalisation} (adopting
minimum wavelet scale $j_0 = 0$ and maximum wavelet scale $j_{\text{max}} = 10$
resulting in a total of 11 wavelet scales).  Additional complexity may produce better results at the
cost of computational efficiency. Furthermore scale-discretised wavelets are only one possible choice of spherical wavelets
(see Section~\ref{sec:Prior_function}).  Other wavelets on the sphere could be adopted and are
interchangable within this reconstruction formalism, provided they support exact synthesis of a signal from its wavelet coefficients.
\par
We adopt the signal to noise ratio (SNR) as a metric to compare how closely each
convergence estimator matches the true convergence map. This recovered SNR in
decibels (dB) is defined to be,
\begin{equation} \label{eq:snr}
\text{Recovered SNR} = 20 \times \log_{10}\Bigg (\frac{\norm{\kappa}_2}{\norm{\kappa - \kappa^{\text{map}}}_2} \Bigg ),
\end{equation}
from which it is clear that the larger the recovered SNR the more accurate\footnote{Accuracy here is in regard to the
pixel-level deviation not structural correlation, for which specific estimators may be designed.}
the convergence estimator. Additionally we record the Pearson correlation coefficient between
recovered convergence estimators $\kappa^{\text{map}} \in \mathbb{C}^{N_{\mathbb{S}^2}}$
and the fiducial convergence $\kappa \in \mathbb{C}^{N_{\mathbb{S}^2}}$ as a measure of
topological fidelty of the estimator. The Pearson correlation coefficient is defined to be
\begin{equation}
r = \frac{ \sum_{i=1}^{N_{\mathbb{S}^2}} \lbrace \kappa^{\text{map}}(i) - \bar{\kappa}^{\text{map}} \rbrace
\lbrace \kappa(i) - \bar{\kappa} \rbrace }{ \sqrt{\sum_{i=1}^{N_{\mathbb{S}^2}} \lbrace \kappa^{\text{map}}(i) -
\bar{\kappa}^{\text{map}} \rbrace^2} \sqrt{\sum_{i=1}^{N_{\mathbb{S}^2}} \lbrace \kappa(i) - \bar{\kappa} \rbrace^2}  },
\end{equation}
where $\bar{x} = \langle x \rangle$. The correlation coefficient $r \in [-1,1]$ quantifies the
structural similarity between two datasets: 1 indicates maximal positive correlation, 0
indicates no correlation, and -1 indicates maximal negative correlation.
\par
In practice the SKS estimator (as with its predecessor the KS estimator) is
post-processed \textit{via} axisymmetric convolution with an often quite large Gaussian
smoothing kernel. The absolute scale of this kernel is typically chosen `by eye' (which is to
say arbitrarily), but in order to maximise the performance of the SKS estimator we
iteratively compute the smoothing scale which maximises the recovered SNR, yielding the
best possible reconstruction that can be provided by the SKS estimator (\textit{i.e.} with optimal
smoothing). We then use this optimal SKS estimator for comparison. Note that this may
only be performed in simulation settings where the fiducial convergence is known. Further
note that such \textit{ad hoc} parameters do not exist within the DarkMapper formalism,
for which a principled statistical problem is posed and solved by automated optimization
algoithms.

\subsubsection{Noise} \label{sec:Noise_computation}
For weak-lensing surveys the noise level of a given pixel is dependent on: the number
density of galaxy observations $n_{\text{gal}}$ (typically given per arcmin$^2$), the size
of said pixel, and the variance of the intrinsic ellipticity distribution $\sigma_e^2$.
\par
Knowing the area $A$ of a given pixel the noise standard deviation $\sigma_i$ is simply given
by,
\begin{equation} \label{eq:noise_variance}
\sigma_i = \sqrt{\frac{\sigma_e^2}{A \times ( 180 / \pi)^2 \times 3600 \times n_{\text{gal}} }},
\end{equation}
where $3600(180 / \pi)^2$ converts steradians to arcmin$^2$ \emdash this relation is
simply a reduction in the noise standard deviation by the root of the number of data-points.
Thus, larger pixels which (assuming a roughly uniform spatial distribution of galaxy
observations) contain more observations have smaller noise variance. In
practice the value of $\sigma_i$ (and therefore the covariance $\Sigma$) can be determined
using the true number of galaxies in a given pixel rather than $n_{\text{gal}}$.
\par
The typical intrinsic ellipticity standard deviation is $\sigma_e \sim 0.37$. Upcoming Stage
\rom{4} surveys (\textit{e.g.} Euclid \citep{EUCLID} and LSST) are projected to achieve a number density
of $n_{\text{gal}} \sim 30$ per arcmin$^2$ \emdash a soft limit due to blending complications.
For academic discussion we also consider the case of a potential future space-based
survey which may push the number density as high as $n_{\text{gal}} \sim 100$ per
arcmin$^2$, in addition to lower number densities $n_{\text{gal}} \in [5, 10]$ per arcmin$^2$
which are representative of past Stage \rom{3} surveys.

\begin{figure*}
{\Large Ground Truth convergence} \par\medskip
  \begin{subfigure}{.54\textwidth}
  \includegraphics[width=0.65\textwidth, right, trim={0 1.0cm 13.4cm 1.7cm},clip]{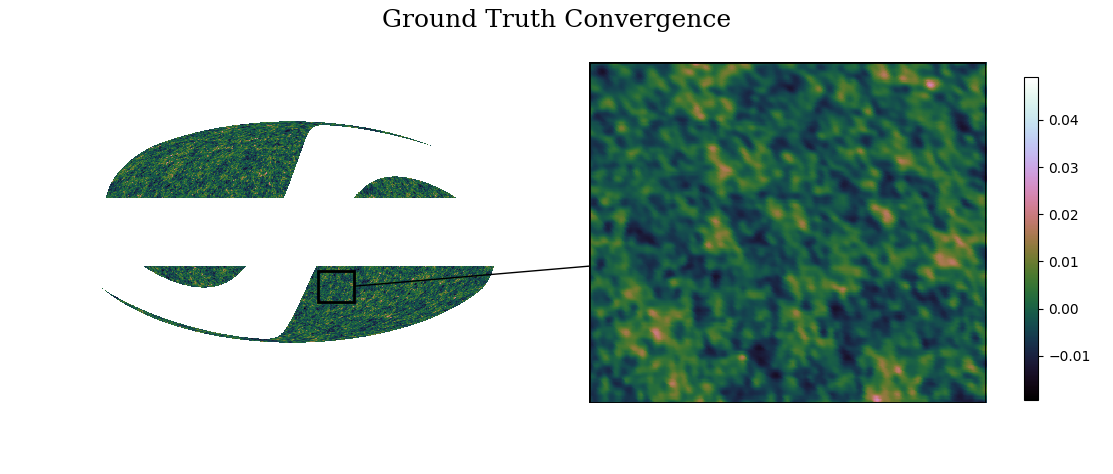}
\end{subfigure}%
\begin{subfigure}{.46\textwidth}
  \centering
  \includegraphics[width=0.52\textwidth, left, trim={14.93cm 1.0cm 0 1.9cm},clip]{Images/Ground_truth_L_2048.png}
\end{subfigure}
  \put(-285,-48){\Large SKS smoothed} \put(-418,-48){\Large SKS} \put(-140,-48){\Large DarkMapper (ours)}  \\
  \includegraphics[width=0.92\textwidth, trim={2.0cm 2.0cm 3.0cm 2.4cm},clip]{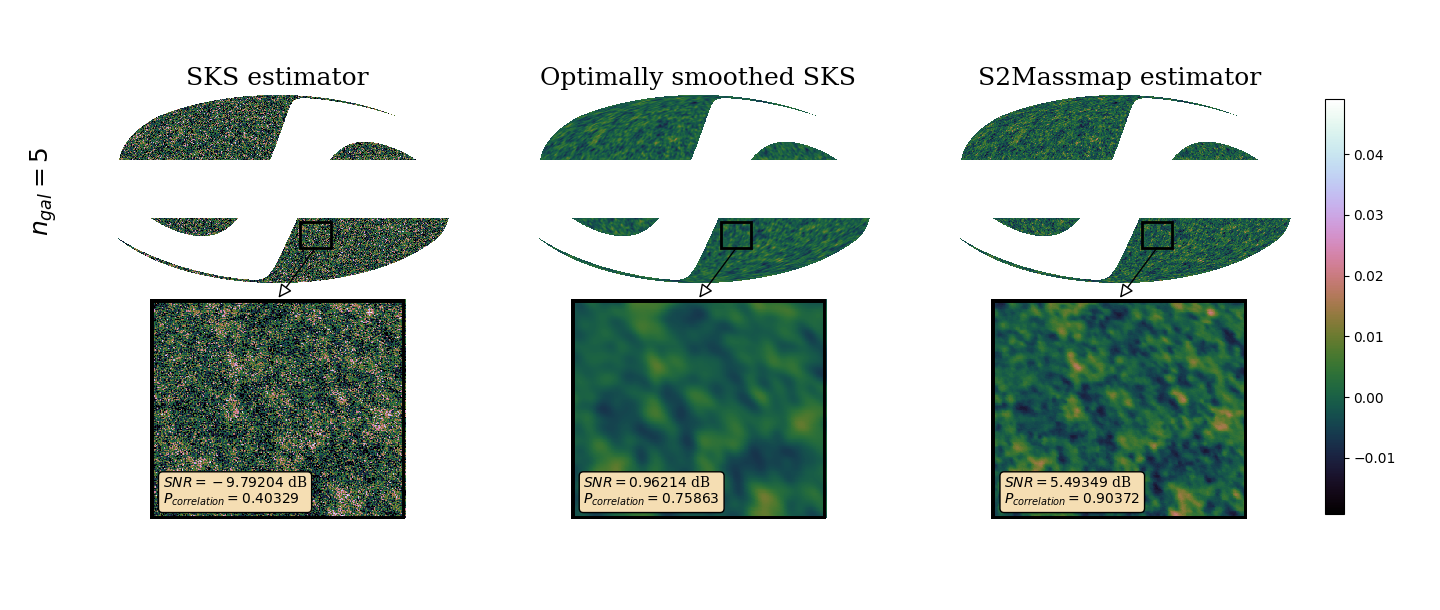}
  \put(-470,80){\large \rotatebox[origin=c]{90}{n$_{\text{gal}} = 5$}} \put(-490,80){\large \rotatebox[origin=c]{90}{Stage \rom{3}}}\\
  \includegraphics[width=0.92\textwidth, trim={2.0cm 2.0cm 3.0cm 2.4cm},clip]{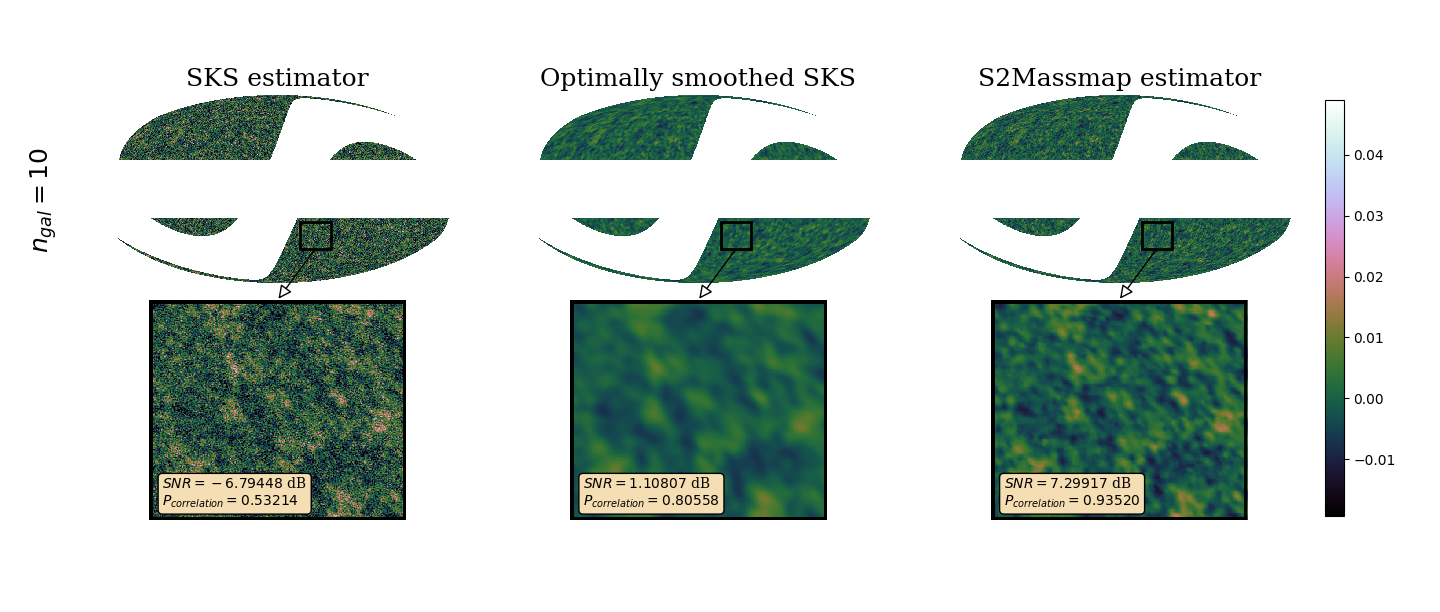}
  \put(-470,80){\large \rotatebox[origin=c]{90}{n$_{\text{gal}} = 10$}}\\
  \includegraphics[width=0.92\textwidth, trim={2.0cm 2.0cm 3.0cm 2.4cm},clip]{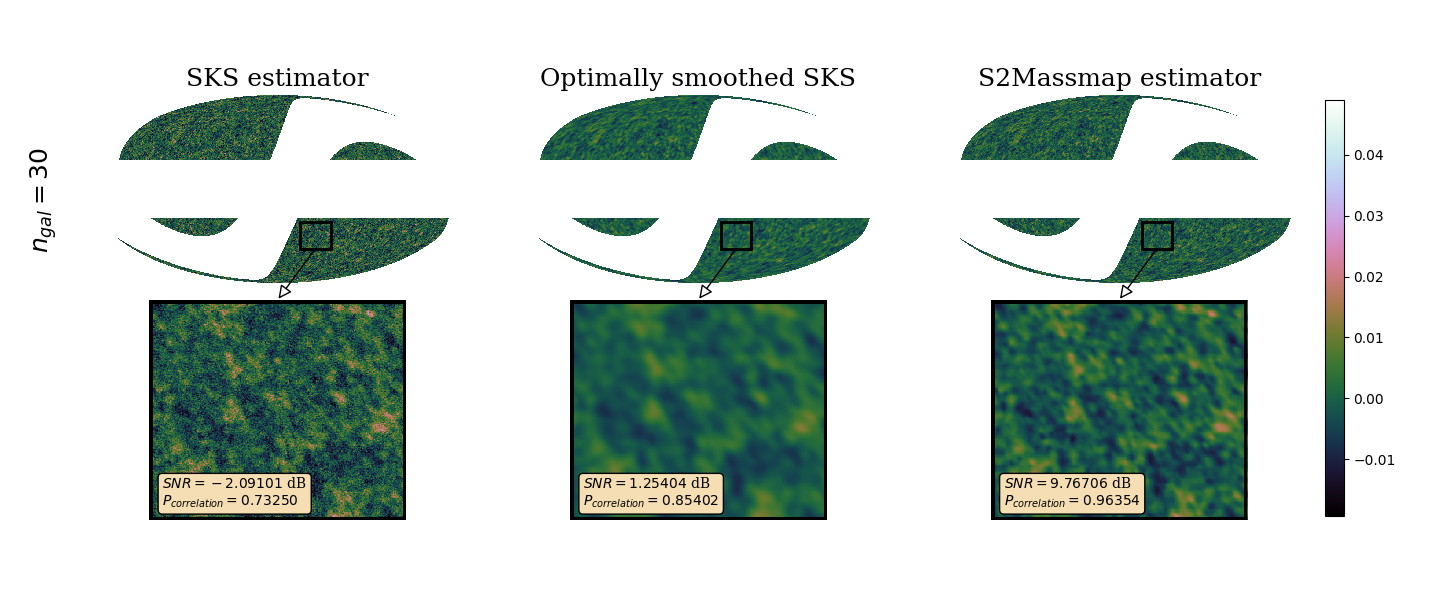}
  \put(-470,80){\large \rotatebox[origin=c]{90}{n$_{\text{gal}} = 30$}} \put(-490,80){\large \rotatebox[origin=c]{90}{Stage \rom{4}}}\\

    \caption{ The top row displays the ground truth Takahashi convergence map as described
    in Section \ref{sec:dataset} with close up of a small region. \textbf{Top to bottom:} Increasing
    number density of galaxies ($n_{\text{gal}}$) and therefore decreasing noise levels. At the
    top we have $n_{\text{gal}} =  5$ which is representative of current Stage \rom{3} surveys,
    at the bottom we have $n_{\text{gal}} = 30$ which has been forecast for upcoming Stage
    \rom{4} surveys (\textit{e.g.} Euclid / LSST). \textbf{Left to right:} The spherical Kaiser-Squires
    \citep{[3]} estimator without the \textit{ad hoc} smoothing kernel postprocessing, the
    optimally smoothed spheircal Kaiser-Squires estimator, and finally  the DarkMapper estimator.
    \textbf{Discussion:} Clearly the DarkMapper estimator is visibly superior in all cases,
    numerically recovering both significantly larger SNR and Pearson correlation coefficients.
    All reconstructions have are plotted on the same colorscale to aid comparison
    \citep{Cubehelix}.
    }
    \label{fig:reconstruction_fidelity}
\end{figure*}

\subsection{Reconstruction results}
For an angular bandlimit $\ell_{\text{max}} = 2048$, a pseudo-Euclid mask and input
$n_{\text{gal}} \in [5, 10, 30, 100]$ we compute the spherical Kaiser-Squires (SKS)
estimator, an idealized (optimally smoothed) SKS estimator, and the DarkMapper estimator.
The results can be found in Figure \ref{fig:reconstruction_fidelity} and numerically in Table
\ref{table:recon_fidelity}. In all cases the DarkMapper estimator provides the highest
reconstruction fidelity both in terms of recovered SNR and Pearson correlation coefficient.
Note that for $\ell_{\text{max}}=2048$ and the number density of galaxy observations selected
the mean number of galaxies per pixel is $\mathcal{O}(10-10^3)$.

It is important to note that the optimal smoothing kernel for the SKS estimator cannot be
known and thus in practice is often selected `by eye' which is to say selected \textit{ad hoc}.
Therefore the smoothed SKS results here constitute an upper bound. The DarkMapper
framework is fully principled and requires no \textit{ad hoc} parameter selection and is therefore
likely to perform in much the same way when applied to observational data.

For Stage \rom{3} survey settings with $n_{\text{gal}} = 5, 10$ the increase in SNR ($\Delta$ SNR)
of the DarkMapper estimator  over the SKS (optimally smoothed SKS) estimator
was $+15.286 \; (+4.532)$ dB and $+14.093 \; (+6.191)$ dB respectively. Recall that dB is measured
on a logarithmic scale (see equation \ref{eq:snr}) and so this increase is quite dramatic. Furthermore
the Pearson correlation coefficient increased from $0.403 \; (0.759)$ to $0.904$ and $0.532 \; (0.860)$
to $0.935$ for $n_{\text{gal}} = 5, 10$ respectively.

For the Stage \rom{4} Euclid-type setting with $n_{\text{gal}} = 30$ $\Delta$, SNR was found
to be $+11.858 \; (+8.513)$ dB, along with which the Pearson correlation coefficient rose from
$0.723 \; (0.854)$ to $0.964$. As this setting is highly representative of the observations which
will be made in Stage \rom{4} surveys this strongly suggests that algorithms such
as DarkMapper should be adopted for weak lensing mass-mapping.

\section{Application to public data} \label{sec:public_data}

Finally we apply both the SKS and DarkMapper estimators to a collated map of the majority
of the public wide field weak lensing observational datasets in order to reconstruct a single
global dark-matter mass-map computed natively on the sphere. Furthermore we demonstrate
straightforward global uncertainty quantification on our reconstruction.

Specifically we perform convergence reconstructions on the DESY1 \citep{DES_ref1, DES_ref2, DES_ref3},
CFHTLens \citep{CFHTLens_ref}, and the KiDS450
\citep{KiDS_Hildebrandt_and_Viola_2017, KiDS_Fenech_2017}
weak lensing shear datasets. See specific acknowledgements and related papers for further
details. Note that throughout we have not chosen to perform reduced shear iterations, assuming
that the observed shear is approximately the reduced shear $\gamma \sim g$ (in a more
detailed analysis one could perform such further iterations)

\begin{figure*}
	\centering
  \includegraphics[width=0.9\textwidth, trim={3.6cm 1.2cm 5.1cm 1.3cm},clip]{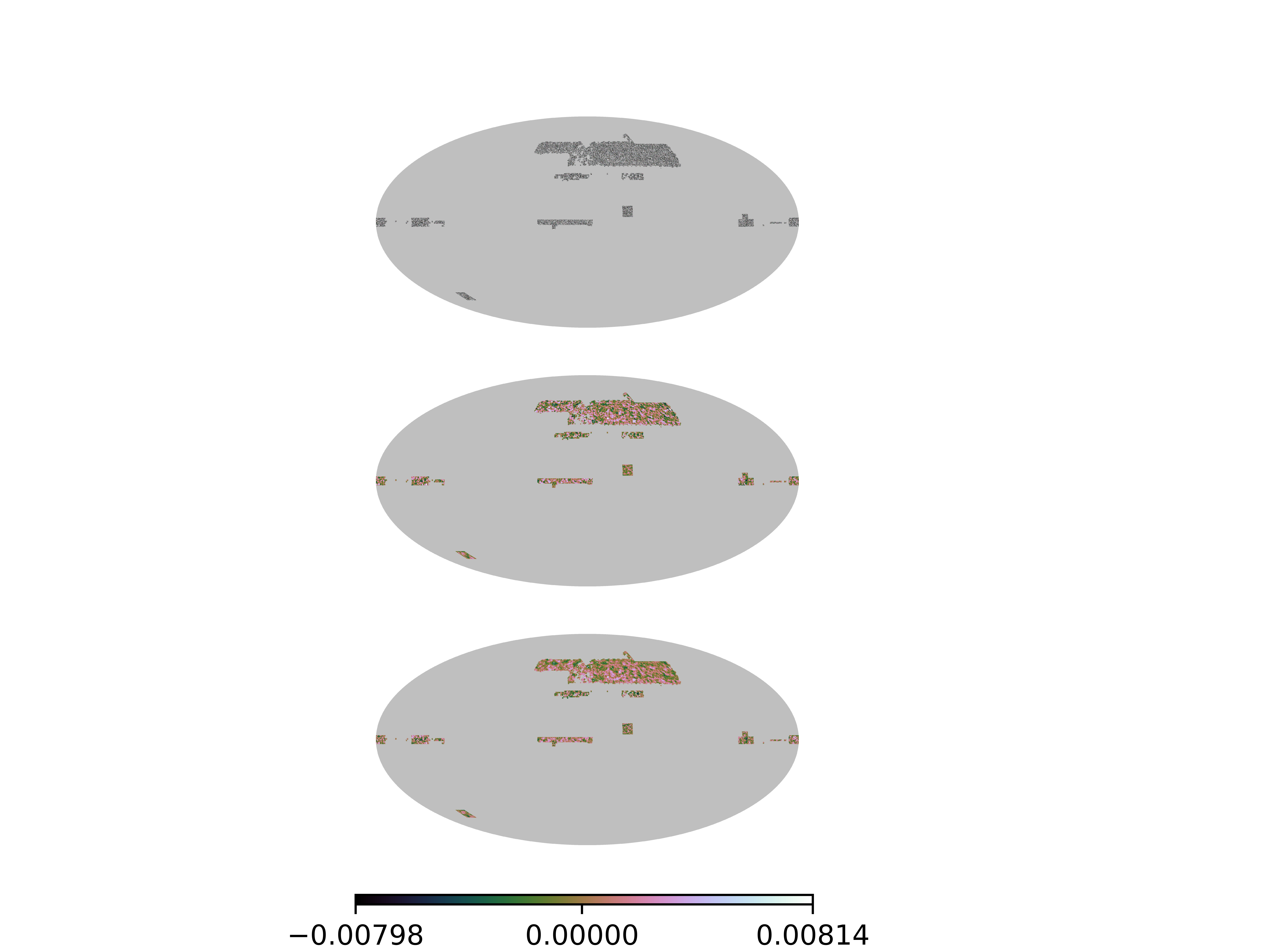}
  \put(-255,183){\Large DarkMapper (ours)} \put(-250,383){\Large SKS smoothed} \put(-220,583){\Large SKS} \\
  \includegraphics[width=0.5\textwidth, trim={3.6cm 0.0cm 5.1cm 11.2cm},clip]{Images/dpi2000_Spherical_Joint_reconstruction.png}
    \caption{
   Mollweide projections of global reconstruction of the majority of public weak lensing datasets.
    \textbf{Top to bottom:} Spherical Kaiser-Squires (SKS) estimator without Gaussian
    smoothing kernel, SKS estimator with FWHM = $\Theta = 25$ arcmin smoothing kernel (as in other studies),
    DarkMapper (our) estimator. All reconstructions are plotted on the same colorscale to aid comparison \citep{Cubehelix}.
    The data-sets can be found online at \url{https://doi.org/10.5281/zenodo.3980652}.
    }
    \label{fig:spherical_joint_reconstruction}
\end{figure*}

\begin{figure*}
	\centering
  \begin{subfigure}{.5\textwidth}
  \includegraphics[width=1.0\textwidth, trim={4.3cm 14.1cm 15cm 2.30cm},clip]{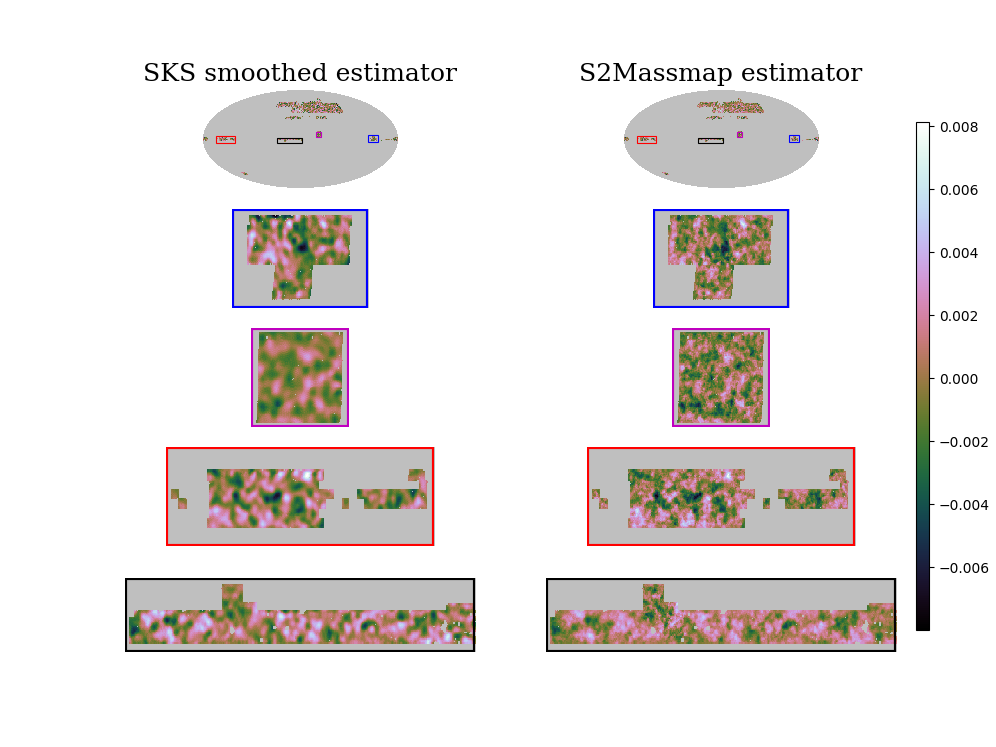}
\end{subfigure}%
\put(-145,65){\Large SKS smoothed}
\begin{subfigure}{.5\textwidth}
  \centering
  \includegraphics[width=1.0\textwidth, trim={14.5cm 14.1cm 4.5cm 2.30cm},clip]{Images/Joint_reconstruction_L_2048.png}
\end{subfigure}
\put(-140,65){\Large DarkMapper (ours)}\\
  \includegraphics[width=0.9\textwidth, trim={1.5cm 3.5cm 1.9cm 6.8cm},clip]{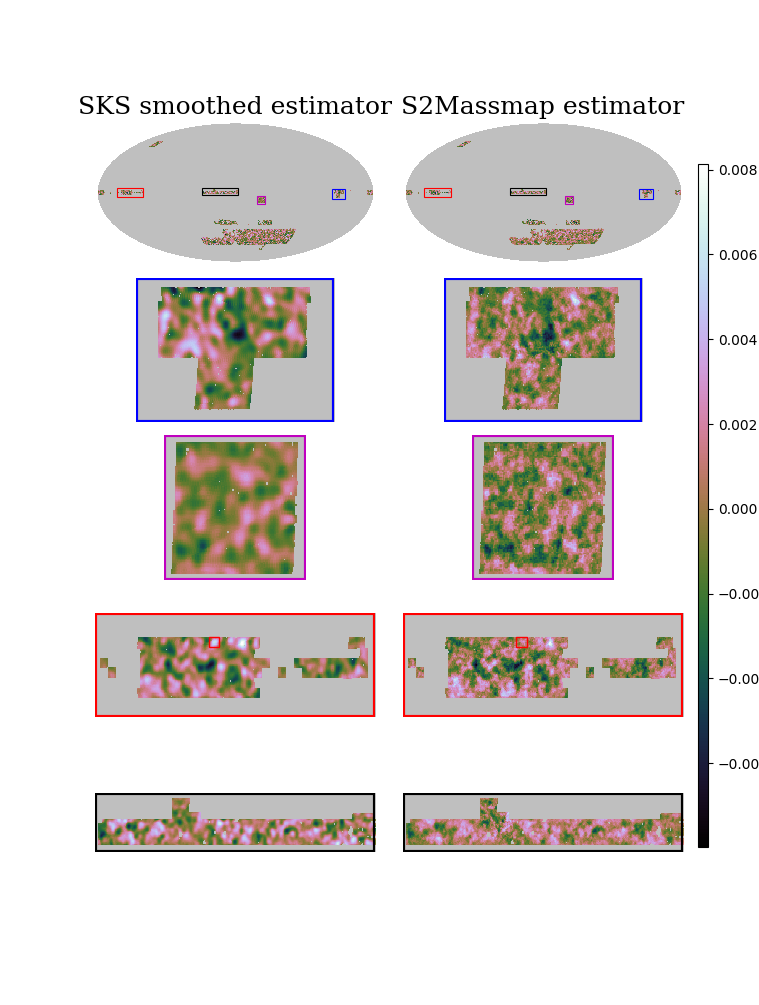}
   \put(-450,370){\Large \rotatebox[origin=c]{90}{KiDS450}}
   \put(-450,260){\Large \rotatebox[origin=c]{90}{CFHTLens}}
   \put(-450, 140){\Large \rotatebox[origin=c]{90}{KiDS450}}
   \put(-450,30){\Large \rotatebox[origin=c]{90}{DES Y1}}  \\
  \includegraphics[width=0.5\textwidth, trim={3.6cm 0.0cm 5.1cm 11.2cm},clip]{Images/dpi2000_Spherical_Joint_reconstruction.png}
    \caption{
    \textbf{Top to bottom:} Global reconstruction of the majority of public weak lensing datasets, magnified
    view of a variety of patches.
    \textbf{Left to right:} Spherical Kaiser-Squires (SKS) estimator without Gaussian
    smoothing kernel, SKS estimator with FWHM = $\Theta = 25$ arcmin smoothing kernel (as in other studies),
    DarkMapper (our) estimator.
    \textbf{Discussion:} Note the overall agreement between both the smoothed SKS
    estimator and our DarkMapper estimator, however notice the significant increase in
    small-scale detail captured by the DarkMapper estimator. All reconstructions are plotted
    on the same colorscale to aid comparison \citep{Cubehelix}. These data-sets can be found
    online at \url{https://doi.org/10.5281/zenodo.3980652}.
    }
    \label{fig:joint_reconstruction}
\end{figure*}

\subsection{Joint Spherical Mass-Map}

All aforementioned weak lensing shear observational datasets were collated into a single
joint global dataset. For each data-set we select only galaxies with non-zero
catalog weight $w(i) > 0$ and perform a correction for the multiplicative bias by $w(i)$ and
additive by $c_{1,2}(i)$  biases per observation. Specifically this correction for ellipticities
$e_{1,2}(i)$ is given by
\begin{align}
  \Re \lbrace \gamma(i) \rbrace = \frac{\sum_j w(j) \lbrace
  e_{1}(j) - c_{1}(j) \rbrace}{ \sum_j w(j) \lbrace 1 + \text{mcorr}(j) \rbrace }, \nonumber \\
  \Im \lbrace \gamma(i) \rbrace = \frac{\sum_j w(j) \lbrace
  e_{2}(j) - c_{2}(j) \rbrace}{ \sum_j w(j) \lbrace 1 + \text{mcorr}(j) \rbrace },
\end{align}
where $j$ are observations such that observation $j$ belongs to pixel $i$, mcorr is the catalog
magnification correction and $\Re, \Im$ denote the real and imaginary components of the
shear field $\bm{\gamma}$ respectively.

This joint global dataset was then projected onto an equiangularly sampled (MW) spherical
shear map $\gamma_{\text{obs}}$ with an angular bandlimit of $\ell_{\text{max}} = 2048$.
During this projection the number of galaxies projected into each pixel was recorded to
create a complimentary map of observations per pixel, from which the data covariance
$\Sigma_{\text{obs}}$ is directly determined (as discussed in Sections \ref{sec:Likelihood_function}
and \ref{sec:Noise_computation})

To this spherical shear map $\gamma_{\text{obs}}$ (with corresponding data covariance
$\Sigma_{\text{obs}}$) we apply DarkMapper outlined in
Section~\ref{sec:bayesian_inference} with the same parameter choices outlined in Section~\ref{sec:Sim_testing} \citep[see][for a planar equivalent]{M1}.
Additionally, we provide the SKS \citep{[3]} reconstruction
which we present in both its fundamental form (without post-processing Gaussian smoothing)
and in its typical form (with post-processing Gaussian smoothing with full width at half
maximum FWHM = $\Theta = 25$ arcmins). All data products aforementioned within this
section are publicly available and may be found at \url{https://doi.org/10.5281/zenodo.3980652}.

The results of all reconstruction algorithms can be seen globally in Figure
\ref{fig:spherical_joint_reconstruction} and with enhanced regions in Figure \ref{fig:joint_reconstruction},
where all subplots share the same colourscale \citep{Cubehelix}. It is immediately apparent
that the SKS estimator, in the absence of smoothing, is overwhelming dominated by noise
(hence the motivation for post-processing).

In contrast to this, the SKS estimator with a $\Theta = 25$ arcmin post-processing Gaussian
smoothing is largely in agreement with the DarkMapper estimator, however this smoothed
SKS estimator unsurprisingly lacks any significant small-scale structure. Further
note that the smoothed SKS estimator does not mirror all high intensity structure (\textit{e.g.}
peaks and voids) recovered by the DarkMapper estimator, which indicates more significant deviations
between the two estimators. The most egregious of these cases is highlighted in the red boxed KiDS450
patch of Figure~\ref{fig:joint_reconstruction}.

These structural dissimilarities between the smoothed SKS and DarkMapper estimator may
reasonably be attributed to large noise fluctuations and boundary effects, both of which are
not reasonably accounted for by the SKS estimator. Observation of such significant differences
indicates that more principled reconstruction algorithms (such as DarkMapper) are important
considerations when attempting to perform future statistical and scientific inference from dark
matter mass-maps.

All reconstructions were performed on a 2016 MacBook air and took $\sim 30$ hours to compute.
A further $\sim 100$ hours were optionally undertaken for annealing iterations to optimise
the convergence. Note that this is by no means a benchmark of computational performance.

\begin{table*}
  	\centering
\begin{tabular}{|p{1.3cm}p{1.3cm}||p{2.0cm}p{2.0cm}||p{2cm}|}
\hline
\multicolumn{2}{|c||}{Surrogate $\kappa^{\text{sur}}$}  & \multicolumn{2}{c||}{Analysis ($\epsilon^{\prime}_{99\%} = 842789$)} & \multicolumn{1}{c|}{Hypothesis test} \\
\hline
Estimator & $\Theta$ (arcmin)  & Obj($\kappa^{\text{sur}}$) & Obj($\kappa^{\text{sur}}) / \epsilon^{\prime}_{99\%}$  & $\kappa^{\text{sur}} \in C^{\prime}_{99\%}$ \\
\hline \hline
\rowcolor{red!20}
SKS  & 0 & 6151070  & 7.298  & \multicolumn{1}{c|}{\xmark} \\
\rowcolor{red!20}
& 5 & 2584510 & 3.067 & \multicolumn{1}{c|}{\xmark}  \\
\rowcolor{red!20}
& 10 & 891266 & 1.058 & \multicolumn{1}{c|}{\xmark}  \\
& 15 & 586741 & 0.696 & \multicolumn{1}{c|}{\cmark}  \\
& 20 & 513223 & 0.609 & \multicolumn{1}{c|}{\cmark} \\
& 25 & 488887 & 0.580 & \multicolumn{1}{c|}{\cmark} \\
 & 30 & 478245 & 0.567 & \multicolumn{1}{c|}{\cmark} \\
\hline
\end{tabular}
\caption{\textbf{Description:} Uncertainty quantification of convergence estimators and
smoothing scales, in each case the convergence surrogate solution $\kappa^{\text{sur}}$
is defined by estimator (\textit{i.e.} SKS) and Gaussian smoothing scale
FWHM = $\Theta$ in arcmins. For each surrogate both the objective function and the ratio
of the objective function to the level set threshold at $99\%$ confidence
$\epsilon^{\prime}_{99\%}$ is presented. The right hand column indicates whether a given
surrogate $\kappa^{\text{sur}}$ belongs to the credible set (and is therefore not rejected as
a possible solution to the reconstruction). Shaded in red are solutions which are rejected by
Bayesian hypothesis testings.
\textbf{Discussion:} Clearly, the SKS estimator without smoothing is unequivically rejected,
which is concurrant with the community's intuition that smoothing is required for the SKS
estimator to produce physically meaningful solutions.
The minimal smoothing scale required for any SKS solution to not be rejected is
$\Theta \sim 15$ arcminutes, therefore with a typical smoothing of $\Theta \in [25, 30]$
arcmins the SKS solution belongs to the DarkMapper credible set and cannot be
rejected at $99\%$ confidence (\textit{i.e.} the two estimators are not necessarily conflicting).  Nevertheless, the DarkMapper estimator contains greater fine-scale structure.
}
\label{table:smoothing_uncertainty}
\end{table*}%

\subsection{Local uncertainty quantification}

Given significant structural dissimilarities between the SKS and DarkMapper convergence
estimators we performed several hypothesis tests of local structure. Specifically we
addressed the missing peaks observed in the smoothed SKS estimator of the lower (red)
region of Figure \ref{fig:joint_reconstruction} but not in the corresponding DarkMapper
estimator. We did so by performing local hypothesis testing of structure as described
in Section \ref{sec:Hypothesis_testing} \citep[see ][ for more comprehensive details]{M1}.

In all cases the hypothesis test of local structure could not reject the existance of such
structure at reasonable confidence. This is unsurprising given the notably high noise
level inherent to Stage \rom{3} weak lensing surveys (which reduces the magnitude of the
objective function thus making the approximate level set threshold $\epsilon_{\alpha}^{\prime}$
more difficult to reach; see Section \ref{sec:UQ}) and the extremely high dimensionality
$\sim \mathcal{O}(10^7)$ of the reconstruction (which directly increases the level set
threshold $\epsilon_{\alpha}^{\prime}$ in equation \ref{eq:level_set_threshold}).

\subsection{Global uncertainty quantification}

For high dimensional cases it is often more informative to consider global
features of the reconstruction, as discussed in Section \ref{sec:UQ}. A question one may
wish to address is for which smoothing scales $\Theta$ does the SKS
estimator provide solutions that are not in disagreement with the DarkMapper estimator
at some well defined confidence.

To address this question within our global uncertainty quantification we consider the SKS
estimator with a variety of Gaussian smoothing kernels, specifically $\Theta = 5i $ for integer $i \in [0,6]$,
which is to say a uniform sampling of different (in practice arbitrary) smoothing choices
ranging from no smoothing (the basic SKS estimator) to the typically adopted case of
$\Theta \sim 30$ arcmin smoothing. In this way we can directly address the question of
which smoothing scales produces solutions $\kappa^{\text{sur}}(\Theta)$ belong to
the DarkMapper approximate HPD-credible region $C_{\alpha}^{\prime}$ (are consistent
with the DarkMapper estimator) and which solutions are unacceptable (\textit{i.e.}
those solutions which reject the null hypothesis that the surrogate is within the credible set)
at $100(1-\alpha)\%$ confidence.

The results of this global uncertainty quantification (at $99\%$ confidence) can be found
numerically in Table \ref{table:smoothing_uncertainty}. Despite the high noise level present
in the joint dataset, the uncertainty quantification technique is sensitive enough to reject the
SKS estimator for $\Theta = 0, 5, 10$ arcminutes, which is to say that these smoothing
scales are in disagreement with the DarkMapper estimator at $99\%$ confidence and
are unlikely to be physically meaningful. This provides statistically rigourous evidence
for the community's intuition that SKS estimators require considerable smoothing to be considered meaningful.

This raises an interesting point worth noting: the SKS estimator (by construction) locates
solutions within $C^\prime_{\alpha}$ which exhibit relatively little small-scale structure,
whereas the DarkMapper estimator locates solutions within $C^\prime_{\alpha}$  which
retain significantly greater small-scale structure. Therefore, though the two solutions do
not disagree at $100(1-\alpha)\%$ confidence, the DarkMapper estimator places relatively
more probability on small-scale structures.

Note that if both estimators provided details of the HPD credible set then a stronger
discussion of the relative cardinality of the intersection of both HPD credible sets could be
used to quantify the level of statistical agreement. However, in this case the SKS
estimator does not support a principled statistical interpretation and so can only justifiably be treated as
a point estimate.

\section{Conclusions} \label{sec:conclusions}

In this paper we have extended the previously presented \citep{M1} sparse Bayesian
reconstruction formalism to the spherical setting, resulting in a sparse spherical Bayesian
mass-mapping algorithm which we refer to as DarkMapper. This algorithm is general
and accomodates any log-concave posterior. In this paper we adopt a Laplace-type
sparsity promoting wavelet prior with a multivariate Gaussian likelihood.

The DarkMapper mass-mapping algorithm was benchmarked against spherical
Kaiser-Squires \citep{[3]} in a variety of realistic weak lensing settings
(ranging from Stage \rom{3} to future space based surveys) using the Takahashi \citep{Ryuichi}
N-body simulations and a pseudo-Euclid masking. In all cases we perform analysis at a
typically adopted angular bandlimit of $\ell_{\text{max}} = 2048$. We do not consider
intrinsic alignments in this paper, but highlight how they may be included should one wish it.

In all simulations the DarkMapper algorithm dramatically outperforms (in both recovered SNR and
recovered Pearson correlation coefficient) the SKS estimator, even when artificially selecting
the optimal SKS smoothing kernel (\textit{i.e.} even when biasing our evaluation in favour
SKS as strongly as possible).

We extend approximate Bayesian uncertainty quantification methods
\citep{[10], M1, M2, M3, [12], Repetti2018} to the spherical setting and explain how one
may leverage these methods from local uncertainty quantification to general global (or aggregate)
uncertainty quantification.

The DarkMapper estimator was applied to a joint observational shear dataset
constructed by collating the majority of publicly available weak lensing data -- specifically the DESY1
\citep{DES_ref1, DES_ref2, DES_ref3}, CFHTLens \citep{CFHTLens_ref}, and the KiDS450
\citep{KiDS_Hildebrandt_and_Viola_2017, KiDS_Fenech_2017}
surveys. To the best of our knowledge this is the first joint spherical reconstruction of all
public weak lensing shear observations.

For comparison we also computed the SKS estimator of this joint dataset. We find, as with
the simulated benchmarking, that the DarkMapper algorithm recovers significantly more
fine-scale structure without the need for any assumptions of Gaussianity or \textit{ad hoc}
smoothing parameters (\textit{i.e.} the smoothing scale for SKS post-processing). This
demonstrates that the algorithm works as expected on observational data.

Finally, uncertainty quantification was carried out to determine for which smoothing scales
the SKS point estimates provide solutions that are acceptable solutions to the DarkMapper Bayesian inference
problem (\textit{i.e.} within the highest posterior density credible region) -- this is to say the
smoothing scales at which both convergence estimates are not conflicting at $99\%$ confidence.
It was found that all SKS reconstructions with smoothing scales below $\sim 15$ arcmin were
rejected at $99\%$ confidence, indicating that significant smoothing is required for agreement
between the SKS and DarkMapper estimators. This reaffirms the community's understanding
that SKS estimators must undergo significant smoothing to recover physically meaningful convergence maps.

Moreover, we demonstrate that the DarkMapper estimator locates permissible solutions with
significantly greater small-scale structure than those which are located by the SKS estimator.
More constraining statistical statements were limited by the inherently high noise level in current
observational shear data.

With the advent of Stage \rom{4} surveys the pixel noise level is projected to drop dramatically
(due to increased galaxy number density), which will inevitably facilitate significantly more
constraining statistical statements. As the DarkMapper estimator not only provides dramatically
increased reconstruction fidelity over the SKS estimator but also supports a principled
Bayesian interpretation, it will be of important use for application to Stage \rom{4}
datasets.

Note that just as we have extended this sparse hierarchical Bayesian mass-mapping formalism
to the sphere ($\mathbb{S}^2$) one can extend it to the ball
($\mathbb{B}^3$) and thus recover similar results for the case of full 3D mass-mapping.
This is a possible avenue for future investigation.

\section*{Acknowledgements}
The authors would like to thank the development teams of S2INV \citep{S2INV}, SOPT
\citep{SOPT, Onose_2016_SOPT, [39], Carrillo_2013_SOPT}\footnote{https://github.com/astro-informatics/sopt},
SSHT \citep{[52], [41]}\footnote{https://astro-informatics.github.io/ssht/}, S2LET
\citep{[52], [54], Leistedt2013, [42], [48],  McEwenS2DWLocalisation}\footnote{http://astro-informatics.github.io/s2let/}, and
SO3 \citep{[48]}\footnote{http://astro-informatics.github.io/so3/} upon which this work is built.
Additionally the authors would like to thank Dr. Peter Taylor for providing the pseudo-Euclid
mask and Dr. Xiaohao Cai for insightful discussion.

MAP is supported by the Science and Technology Facilities Council (STFC).
TDK is supported by a Royal Society University Research Fellowship (URF).  This work
was also supported by the Engineering and Physical Sciences Research Council (EPSRC)
through grant EP/M0110891 and by the Leverhulme Trust. The Dunlap Institute is funded
through an endowment established by the David Dunlap family and the University of Toronto.

\subsection{DES acknowledgements}
This project used public archival data from the Dark Energy Survey (DES). Funding for the DES Projects has been provided by the U.S. Department of Energy, the U.S. National Science Foundation, the Ministry of Science and Education of Spain, the Science and Technology FacilitiesCouncil of the United Kingdom, the Higher Education Funding Council for England, the National Center for Supercomputing Applications at the University of Illinois at Urbana-Champaign, the Kavli Institute of Cosmological Physics at the University of Chicago, the Center for Cosmology and Astro-Particle Physics at the Ohio State University, the Mitchell Institute for Fundamental Physics and Astronomy at Texas A\&M University, Financiadora de Estudos e Projetos, Funda{\c c}{\~a}o Carlos Chagas Filho de Amparo {\`a} Pesquisa do Estado do Rio de Janeiro, Conselho Nacional de Desenvolvimento Cient{\'i}fico e Tecnol{\'o}gico and the Minist{\'e}rio da Ci{\^e}ncia, Tecnologia e Inova{\c c}{\~a}o, the Deutsche Forschungsgemeinschaft, and the Collaborating Institutions in the Dark Energy Survey.

The Collaborating Institutions are Argonne National Laboratory, the University of California at Santa Cruz, the University of Cambridge, Centro de Investigaciones Energ{\'e}ticas, Medioambientales y Tecnol{\'o}gicas-Madrid, the University of Chicago, University College London, the DES-Brazil Consortium, the University of Edinburgh, the Eidgen{\"o}ssische Technische Hochschule (ETH) Z{\"u}rich,  Fermi National Accelerator Laboratory, the University of Illinois at Urbana-Champaign, the Institut de Ci{\`e}ncies de l'Espai (IEEC/CSIC), the Institut de F{\'i}sica d'Altes Energies, Lawrence Berkeley National Laboratory, the Ludwig-Maximilians Universit{\"a}t M{\"u}nchen and the associated Excellence Cluster Universe, the University of Michigan, the National Optical Astronomy Observatory, the University of Nottingham, The Ohio State University, the OzDES Membership Consortium, the University of Pennsylvania, the University of Portsmouth, SLAC National Accelerator Laboratory, Stanford University, the University of Sussex, and Texas A\&M University.

Based in part on observations at Cerro Tololo Inter-American Observatory, National Optical Astronomy Observatory, which is operated by the Association of Universities for Research in Astronomy (AURA) under a cooperative agreement with the National Science Foundation.

\subsection{Kilo Degree Survey Acknowledgements}
Based on data products from observations made with ESO Telescopes at the La Silla Paranal Observatory under programme IDs 177.A-3016, 177.A-3017 and 177.A-3018.

We use cosmic shear measurements from the Kilo-Degree Survey \citep{KiDS_Kuijken_2015, KiDS_Hildebrandt_and_Viola_2017, KiDS_Fenech_2017}, hereafter referred to as KiDS. The KiDS data are processed by THELI \citep{KiDS_Erben_2013} and Astro-WISE \citep{KiDS_astrowise_2013, KiDS_dejong_2015}. Shears are measured using lensfit \citep{Miller_LensFit_2013}, and photometric redshifts are obtained from PSF-matched photometry and calibrated
using external overlapping spectroscopic surveys \citep[see ][]{KiDS_Hildebrandt_and_Viola_2017}.

\section{Data Availability}

All observational data utilized throughout this paper is publicly available and can be found in the
corresponding references. All joint reconstruction data-sets are publicly available and can be
found online at \url{https://doi.org/10.5281/zenodo.3980652}.
The
DarkMapper reconstruction software will be publicly released in the near future.



\bibliographystyle{mnras}
\bibliography{Refs/references.bib}


\appendix

\section{Wavelets on the Sphere} \label{sec:spherical_wavelets}
This section will provide extremely brief (and quite technical) overview of
scale-discretised spherical wavelets; for extensive details see the related articles
\citep{wiaux2008,[40],[42],[41],[47],[48],[49],[54],McEwenS2DWLocalisation}.
\par
The scale-discretized wavelet transform on the sphere is given by the directional convolution
of each wavelet $\Psi^j$ of scale $j \in [J_0, J_{\text{max}}]$ with a field ${}_0\kappa$,
such that the wavelet coefficients $w^j(\rho) \in \ell_{\text{max}}^2(\text{SO}(3))$ are given by,
\begin{equation}
w^j(\rho) = \int_{\mathbb{S}^2} d\rho(\rho^{\prime}) \; {}_0\kappa(\rho^{\prime})( \mathcal{R}_{\rho} \Psi^j)^*(\rho^{\prime}).
\end{equation}
Additionally, the low frequency component of ${}_0\kappa$ is encapsulated by the
axisymmetric convolution with $\Upsilon \in \ell_{\text{max}}^2(\mathbb{S}^2)$ such that,
\begin{equation}
s(\omega) = \int_{\mathbb{S}^2} d\Omega(\omega^{\prime}) \; {}_0\kappa(\omega^{\prime})(\mathcal{R}_{\omega}\Upsilon)^*(\omega^{\prime}).
\end{equation}
It is possible to project these directional and axisymmetric convolution operators into
harmonic space \citep{[40]},
\begin{equation} \label{eq:harmonic_directional}
(w^j)^{\ell}_{mn} \equiv \langle w^j, \mathcal{D}^{\ell *}_{mn} \rangle =  \frac{8 \pi^2}{2 \ell + 1} \; {}_0\kappa_{lm}\; \Psi^{j*}_{\ell n},
\end{equation}
\begin{equation}\label{eq:harmonic_scalar}
s_{\ell m} \equiv \langle s, Y_{\ell m} \rangle = \sqrt{ \frac{4 \pi}{2 \ell + 1} } \; {}_0\kappa_{lm} \; \Upsilon^*_{\ell 0}.
\end{equation}
The coefficients of a general spherical signal ${}_sf$ can synthesised exactly by
\begin{equation}
  {}_sf(\omega) = \int_{\mathbb{S}^2} d\Omega s(\omega^\prime) (\mathcal{R}_{\omega^\prime} s)(\omega^\prime)
  + \sum_{j=J_0}^{J} \int_{\text{SO}(3)} d\rho w^j(\rho^\prime) (\mathcal{R}_{\rho^\prime} w^j)(\rho^\prime),
\end{equation}
from which we can define the discretized form
of ${}_0\kappa$ by
\begin{align}
{}_0\kappa(\omega) = \sum^{\infty}_{\ell = 0} \sum^{\ell}_{m=-\ell} \Bigg \lbrace & \sqrt{ \frac{4 \pi}{2 \ell + 1} } s_{\ell m} \Upsilon_{\ell 0} Y_{\ell m}(\omega) \nonumber \\
&+ \frac{2 \ell + 1}{8 \pi^2} \sum_{j=J_0}^{J_{\text{max}}} \sum_{n=-\ell}^{\ell} (w^j)^{\ell}_{mn} \; \Psi^{j}_{\ell n} Y_{\ell m}(\omega)  \Bigg \rbrace,
\end{align}
\par
where the harmonic coefficients ${}_0\kappa_{\ell m}$ are explicitly given by
\begin{equation}
{}_0\kappa_{\ell m} = \sqrt{ \frac{2 \ell + 1}{4 \pi} } s_{\ell m} \Upsilon_{\ell 0} + \frac{2 \ell + 1}{8 \pi^2} \sum_{j=J_0}^{J_{\text{max}}} \sum_{n=-\ell}^{\ell} (w^j)^{\ell}_{mn} \; \Psi^{j}_{\ell n}
\end{equation}
Combining equations (\ref{eq:harmonic_directional}) and (\ref{eq:harmonic_scalar}) the
harmonic representation of the spherical wavelet transform at scale $j$ can be written as
a combination of the following linear operators \citep{[40]}:
\begin{equation*}
\hat{w}^j = \bm{\mathsf{N}}^j\bm{\mathsf{W}}^j \; {}_0\hat{\kappa}_{\ell m},
\end{equation*}
are the Wigner coefficients of the directional wavelet coefficients. Here $\bm{\mathsf{W}}^j$
is harmonic space multiplication by the wavelet kernel $\Psi^{j*}_{\ell n}$, and
$\bm{\mathsf{N}}^j$ corresponds to wavelet normalization.
\begin{equation*}
\hat{s} = \bm{\mathsf{S}}\; {}_0\hat{\kappa}_{\ell m},
\end{equation*}
are the spherical harmonic coefficients of the scaling wavelet coefficients, where
$\bm{\mathsf{S}}$ is harmonic space multiplication by  scaling kernel $\Upsilon^*_{\ell 0}$.
\par
We further specialize by grouping the harmonic representation of the coefficients into
$
\hat{\alpha} = \bm{\mathsf{N}} \bm{\mathsf{W}} \; {}_0\hat{\kappa}_{\ell m},
$
where the wavelet scale normalization terms are collectively
$
\bm{\mathsf{N}} = \text{diag}(\mathbb{I}, \bm{\mathsf{N}}^{J_0},\ldots, \bm{\mathsf{N}}^{J_{\text{max}}}),
$
and the Wigner/harmonic space wavelet convolutions are collectively
$
\bm{\mathsf{W}} = \text{diag}(\bm{\mathsf{S}}, \bm{\mathsf{W}}^{J_0},\ldots, \bm{\mathsf{W}}^{J_{\text{max}}}).
$
Finally we introduce a diagonal array of operators $\bm{\mathsf{H}}$ corresponding to
inverse spherical harmonic and Wigner transforms,
$
\bm{\mathsf{H}} = \text{diag}(\bm{\mathsf{Y}}, \bm{\mathsf{\mathcal{D}}}^{J_0},\ldots, \bm{\mathsf{\mathcal{D}}}^{J_{\text{max}}})
$
In the scope of this paper we restrict ourselves to the analysis setting, thus these are the
only operators required though a complete set are provided in \citet{[40]}. Concatenating
these operators we can finally form analysis forward and adjoint wavelet transforms,
\begin{equation} \label{eq:wavelet_transforms}
\tilde{\Psi} = \bm{\mathsf{H}}\bm{\mathsf{N}}\bm{\mathsf{W}}\tilde{\bm{\mathsf{Y}}} \qquad \text{and} \qquad \tilde{\Psi}^{\dagger} = \tilde{\bm{\mathsf{Y}}}^{\dagger}\bm{\mathsf{W}}^{\dagger}\bm{\mathsf{N}}\bm{\mathsf{H}}^{\dagger}.
\end{equation}
Note that $\dagger$ is the adjoint (or conjugate transpose) operator, and as $\bm{\mathsf{N}}$
is simply wavelet scaling it is trivially self-adjoint $\bm{\mathsf{N}} = \bm{\mathsf{N}}^{\dagger}$.
Crucially as we work in the discrete setting the inverse wavelet transforms are not
equivalent to the adjoint wavelet transforms\footnote{An approximation often made throughout
the literature.}, and so we adopt the optimized algorithms to compute fast Wigner and Wigner
adjoint functions presented in the appendix of \citet{[42]}.

\bsp	
\label{lastpage}
\end{document}